# On the use of deep learning for phase recovery

Kaiqiang Wang[1,2,3,*], Li Song[1], Chutian Wang[1], Zhenbo Ren[2], Guangyuan Zhao[3], Jiazhen Dou[4], Jianglei Di[4], George Barbastathis[5], Renjie Zhou[3], Jianlin Zhao[2,*], and Edmund Y. Lam[1,*]

[1]Department of Electrical and Electronic Engineering, The University of Hong Kong, Hong Kong SAR, China
[2]School of Physical Science and Technology, Northwestern Polytechnical University, Xi'an, China
[3]Department of Biomedical Engineering, The Chinese University of Hong Kong, Hong Kong SAR, China
[4]School of Information Engineering, Guangdong University of Technology, Guangzhou, China
[5]Department of Mechanical Engineering, Massachusetts Institute of Technology, Cambridge, Massachusetts, USA

[*]Correspondence: Kaiqiang Wang (kqwang.optics@gmail.com) or Jianlin Zhao (jlzhao@nwpu.edu.cn) or Edmund Y. Lam (elam@eee.hku.hk)

## Abstract

Phase recovery (PR) refers to calculating the phase of the light field from its intensity measurements. As exemplified from quantitative phase imaging and coherent diffraction imaging to adaptive optics, PR is essential for reconstructing the refractive index distribution or topography of an object and correcting the aberration of an imaging system. In recent years, deep learning (DL), often implemented through deep neural networks, has provided unprecedented support for computational imaging, leading to more efficient solutions for various PR problems. In this review, we first briefly introduce conventional methods for PR. Then, we review how DL provides support for PR from the following three stages, namely, pre-processing, in-processing, and post-processing. We also review how DL is used in phase image processing. Finally, we summarize the work in DL for PR and outlook on how to better use DL to improve the reliability and efficiency in PR. Furthermore, we present a live-updating resource (https://github.com/kqwang/phase-recovery) for readers to learn more about PR.

# 1. Introduction

Light, as a complex electromagnetic field, has two essential components: amplitude and phase[1]. Optical detectors, usually relying on photon-to-electron conversion (such as charge-coupled device sensors and the human eye), measure the intensity that is proportional to the square of the amplitude of the light field, which in turn relates to the transmittance or reflectance distribution of the sample (Fig. 1a and Fig. 1b). However, they cannot capture the phase of the light field because of their limited sampling frequency[2].

Actually, in many application scenarios, the phase rather than the amplitude of the light field carries the primary information of the samples[3–6]. For quantitative structural determination of transparent and weakly scattering samples[3] (Fig. 1c), the phase delay is proportional to the sample's thickness or refractive index (RI) distribution, which is critically important for bioimaging because most living cells are transparent. For quantitative characterization of the aberrated wavefront[5] (Fig. 1d and Fig. 1e), the phase aberration is caused by atmospheric turbulence with an inhomogeneous RI distribution in the light path, which is mainly used in adaptive aberration correction. Also, for quantitative measurement of the surface profile[6] (Fig. 1f), the phase delay is proportional to the surface height of the sample, which is very useful in material inspection.

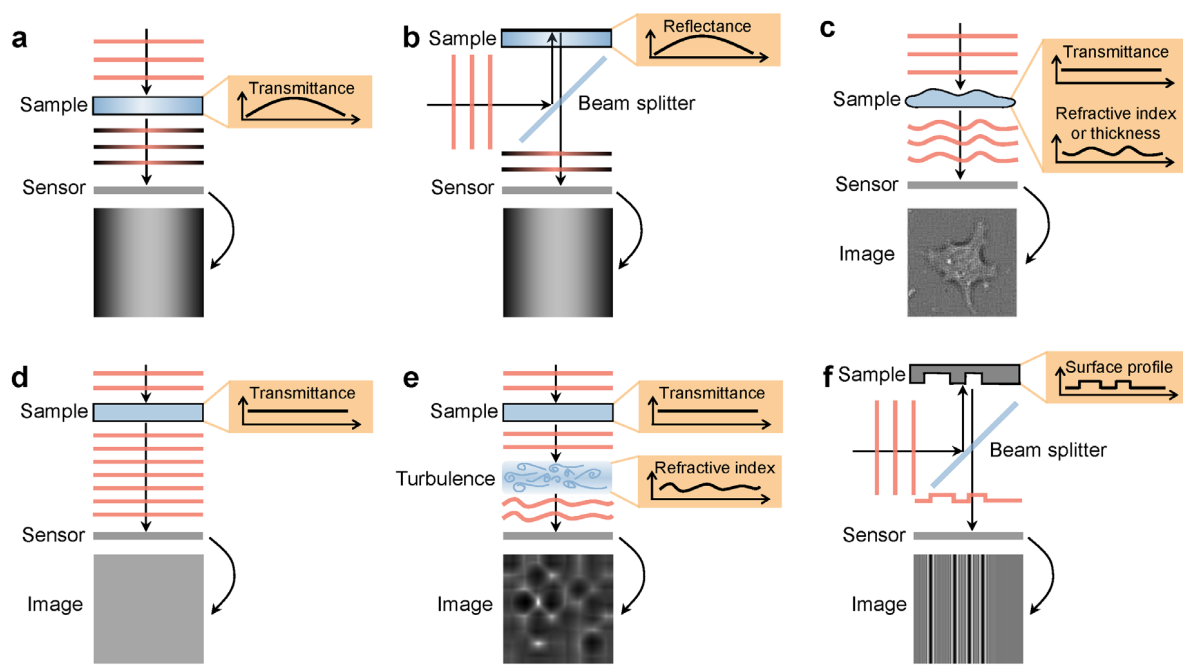

Fig. 1 Light is transmitted through or reflected from different samples. **a** An absorptive sample with a nonuniform transmittance distribution. **b** A reflective sample with a nonuniform reflectance distribution. **c** A transparent (weakly-absorbing) sample with a nonuniform RI or thickness distribution. **d** A sample with a uniform transmittance distribution. **e** A sample with a uniform transmittance distribution placed before atmospheric turbulence with inhomogeneous RI distribution. **f** A reflective sample with a nonuniform surface height distribution.

Since the phase delay across the wavefront is necessary for the above applications, but the optical detection devices can only perceive and record the amplitude of the light field, how can we recover the desired phase? Fortunately, as the light field propagates, the phase delay also causes changes in the amplitude distribution; therefore, we can record the amplitude of the propagated light field and then calculate the corresponding phase. This operation generally comes under different names according to the application domain; for example, it is quantitative phase imaging (QPI) in biomedicine[3], phase retrieval in coherent diffraction imaging (CDI)[4] which is the most commonly used term in x-ray optics and non-optical analogues such as electrons and other particles, and wavefront sensing in adaptive optics (AO)[5] for astronomy and optical communications. Here, we collectively refer to the way of *calculating the phase of a light field from its intensity measurements* as phase recovery (PR).

As is common in inverse problems, calculating the phase directly from an intensity measurement after propagation is usually ill-posed[7]. Suppose the complex field at the sensor plane is known. We can directly calculate the complex field at the sample plane using numerical propagation[8] (Fig. 2a). However, in reality, the sensor only records the intensity but loses the phase, and, moreover, it is necessarily sampled by pixels of finite area size; because of these complications, the complex field distribution at the sample plane generally cannot be calculated in a straightforward manner (Fig. 2b).

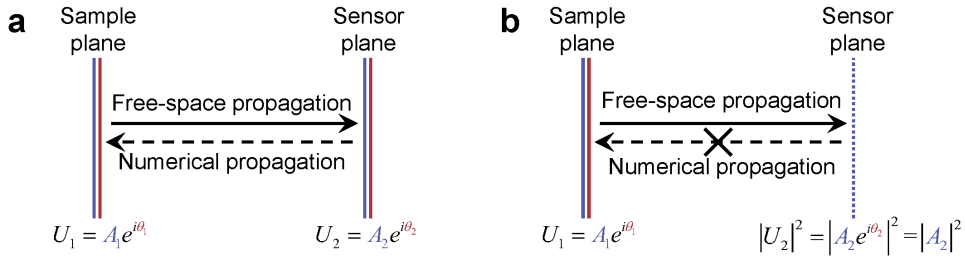

**Fig. 2 Calculating complex field at the sample plane from (a) the complex field or (b) the intensity at the sensor plane.** *U*: complex field. *A*: amplitude. *θ*: phase.

We can transform phase recovery into a well-posed/deterministic problem by introducing extra information, such as holography or interferometry at the expense of having to introduce a reference wave[8,9], transport of intensity equation requiring multiple through-focus amplitudes[10,11], and Shack-Hartmann wavefront sensing which introduces a micro-lens array at the conjugate plane[12,13]. Alternatively, we can solve this ill-posed phase recovery problem in an iterative manner by optimization, i.e., the so-called phase retrieval such as Gerchberg-Saxton-Fienup algorithm[14–16], ptychographic iterative engine[17,18], and Fourier ptychography[19]. Next, we introduce these classical phase recovery methods in more detail.

**Holography/interferometry.** By interfering the unknown wavefront with a known reference wave, the phase difference between the object wave and the reference wave is converted into the intensity of the resulting hologram/interferogram due to alternating constructive and destructive interference of the two waves across their fronts. This enables direct calculation of the phase from the hologram[8].

In in-line holography, where the object beam and the reference beam are along the same optical axis, four-step phase-shifting algorithm is commonly used for phase recovery (Fig. 3)[20]. At first, the complex field of the object wave at the sensor plane is calculated from the four phase-shifting holograms. Next, the complex field at the sample plane is obtained through numerical propagation. Then, by applying the arctangent function over the final complex field, a phase map in the range of $(-\pi, \pi]$ is obtained, i.e., the so-called wrapped phase. The final sample phase is obtained after phase unwrapping. Other multiple-step phase-shifting algorithms are also possible for phase recovery[21]. Spatial light interference microscopy (SLIM), as a well-known QPI method, combines the phase-shifting algorithm with a phase contrast microscopy for phase recovery over transparent samples[22].

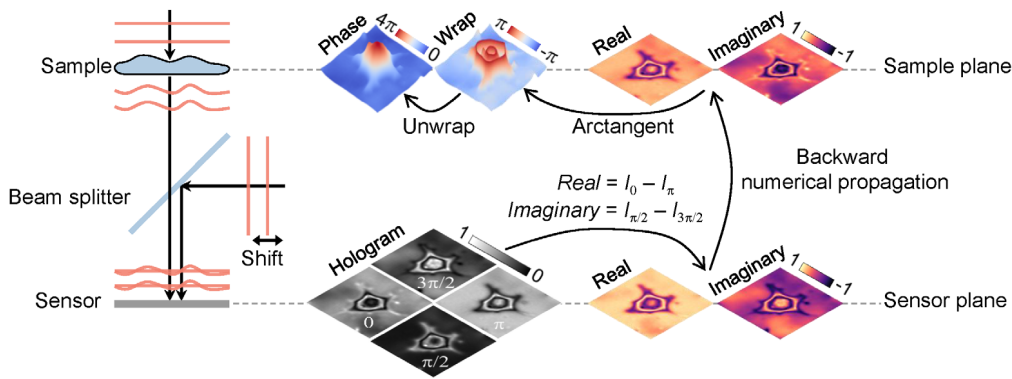

**Fig. 3 Description of in-line holography based on the four-step phase-shifting algorithm.**

In off-axis holography, where the reference beam is slightly tilted from the optical axis, the phase is modulated into a carrier frequency that can be recovered through spatial spectral filtering with only one holographic measurement (Fig. 4)[23]. By appropriately designing the carrier frequency, one can well separate the baseband that contains the reference beam from the object beam. After transforming the measured hologram into the spatial frequency domain through a Fourier transform (FT), we can select the +1st or -1st order beam and move it to the baseband. By applying an inverse FT, the complex sample beam can be retrieved. One has to be careful, however, not to exceed the Nyquist limit on the camera as the angle between reference and object increases. Moreover, as only a small part of the spatial spectrum is taken for phase recovery, off-axis holography typically wastes a lot of spatial bandwidth product of the system. To enhance the utilization of the spatial bandwidth product, the Kramers-Kronig relationship and other iterative algorithms have been recently applied in off-axis holography[24–26].

Both the in-line and off-axis holography discussed above are lensless, where the sensor and sample planes are not mutually conjugate. Therefore a backward numerical propagation from the former to the latter is necessary. The process of numerical propagation can be omitted if additional imaging components are added to conjugate the sensor plane and the sample plane, such as digital holographic microscopy[27].

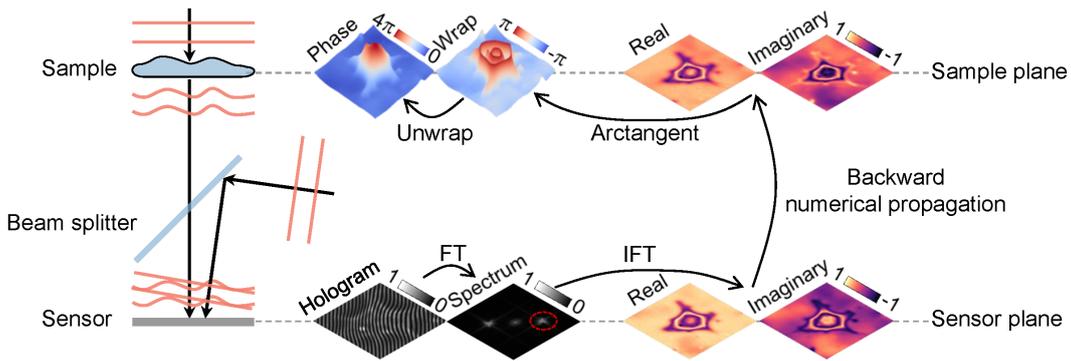

**Fig. 4 Description of off-axis holography based on spatial spectral filtering.**

**Transport of Intensity Equation.** For a light field, the wavefront determines the axial variation of the intensity in the direction of propagation. Specifically, there is a quantitative relationship between the gradient and curvature of the phase and the axial differentiation of intensity, the so-called transport of intensity equation (TIE)[10]. This relationship has an elegant analogy to fluid mechanics, approximating the light intensity as the density of a compressible fluid and the phase gradient as the lateral pressure field. TIE may be derived from the Fresnel-Schrödinger[10], and it is subject to the scalar, paraxial, and weak-defocusing approximations[28,29]. The gradient and curvature of the phase together determine the shape of the wavefront, whose normal vector is then parallel to the wavevector at each point of the wavefront, and consequently to the direction of energy propagation. In turn, variations in the lateral energy flux also result in axial variations of the intensity. Convergence of light by a convex lens is an intuitive example (Fig. 5): the wavefront in front of the convex lens is a plane, whose wavevector is parallel to the direction of propagation. As such, the intensity distribution on different planes is constant, that is, the axial variation of the intensity is equal to zero. Then, the convex lens changes the wavefront so that all wavevectors are directed to the focal point, and therefore as the light propagates, the intensity distribution becomes denser and denser, meaning that the intensity varies in the axial direction (equivalent, its axial derivative is not zero).

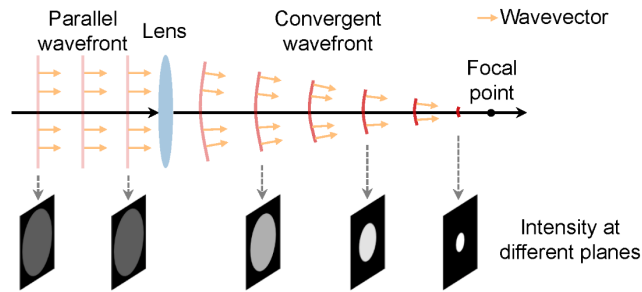

Fig. 5 A convex lens converges light to a focal point.

As there is a quantitative relationship between the gradient and curvature of the phase and the axial differentiation of intensity, we can exploit it for phase recovery (Fig. 6). By shifting the sensor axially, intensity maps at different defocus distances are recorded, which can be used to approximate the axial differential by numerical difference, and thus calculate the phase through TIE. Due to the addition of the imager, the sensor plane and the sample plane are conjugated.

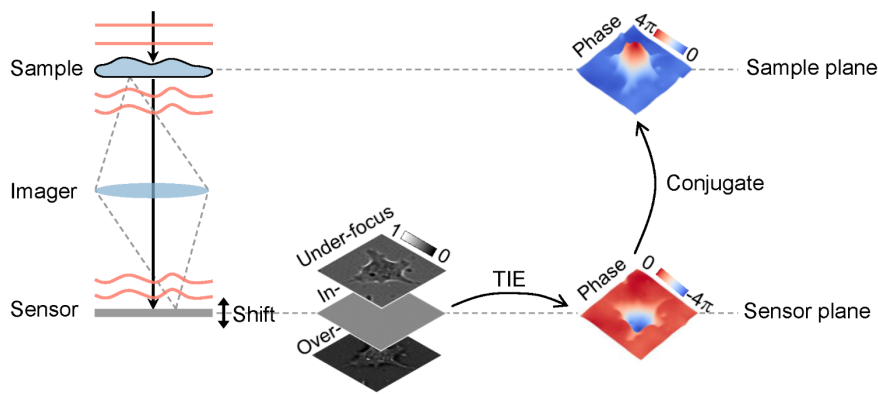

Fig. 6 Description of phase recovery by transport of intensity equation (TIE).

It is worth noting that TIE is suitable for a complete and partially coherent light source, and the resulting phase is continuous and does not require phase unwrapping, while it is only effective in the case of paraxial approximation and weak defocus[11].

**Shack-Hartmann wavefront sensing.** If we can obtain the horizontal and vertical phase gradients of a wavefront in some ways, then the phase can be recovered by integrating the phase gradients in these orthogonal directions. Shack-Hartmann wavefront sensor[12,13] is a classic way to do so from the perspective of geometric optics. It usually consists of a microlens array and an image sensor located at its focal plane (Fig. 7). The phase gradient of the wavefront at the surface of each microlens is calculated linearly from the displacement of the focal point on the focal plane, in both horizontal and vertical ($x$-axis and $y$-axis) directions. The phase can then be computed by integrating the gradient at each point, whose resolution depends on the density of the microlens array. In addition, quantitative differential interference contrast microscopy[30], quantitative differential phase contrast microscopy[31], and quadriwave lateral shearing interferometry[32] also recover the phase from its gradients. There may achieve higher resolution than the Shack-Hartmann wavefront sensor.

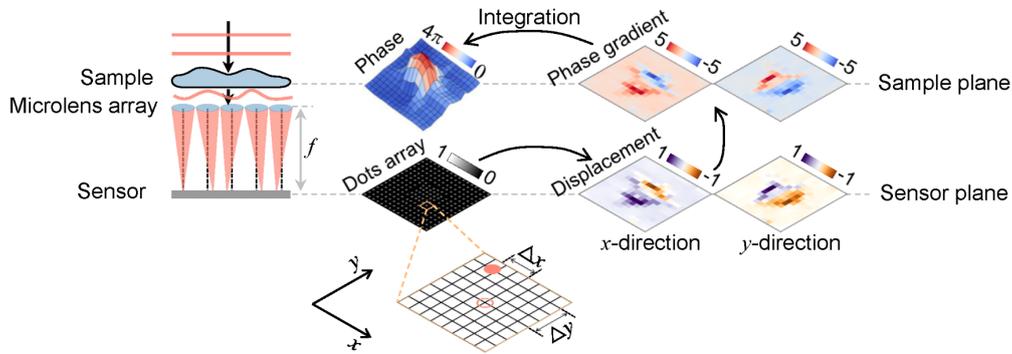

Fig. 7 Description of the Shack-Hartmann wavefront sensor.

**Phase retrieval.** If extra information is not desired to be introduced, then calculating the phase directly from a propagated intensity measurement is an ill-posed problem. We can overcome such difficulty through incorporating prior knowledge. This is also known as regularization. In the Gerchberg-Saxton (GS) algorithm[14], the intensity at the sample plane and the far-field sensor plane recorded by the sensor are used as constraints. A complex field is projected forward and backward between these two planes using the Fourier transform and constrained by the intensity iteratively; the resulting complex field will gradually approach a solution (Fig. 8a). Fienup changed the intensity constraint at the sample plane to the aperture (support region) constraint, so that the sensor only needs to record one intensity map, resulting in the error reduction (ER) algorithm and the hybrid input-output (HIO) algorithm (Fig. 8b)[15,16].

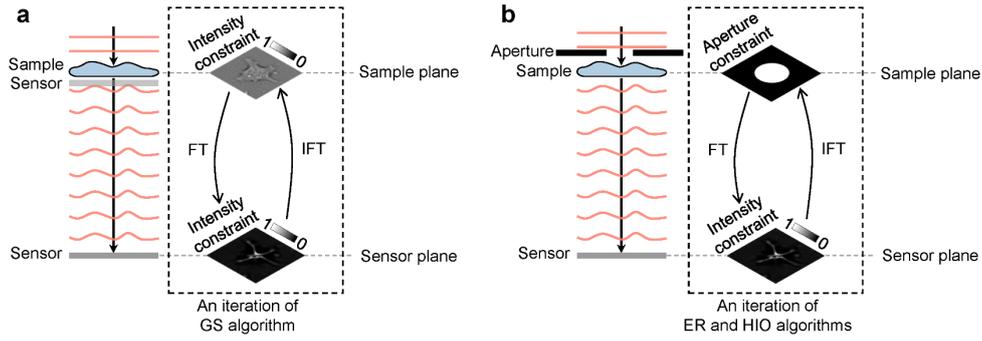

**Fig. 8 Description of alternating-projection algorithms. a** Gerchberg-Saxton algorithm. **b** Error reduction and hybrid input-output algorithms.

Naturally, if more intensity maps are recorded by the sensor, there will be more prior knowledge for regularization, further reducing the ill-posedness of the problem. By moving the sensor axially, the intensity maps of different defocus distances are recorded as an intensity constraint, and then the complex field is computed iteratively like the GS algorithm (Fig. 9a)[33–35]. In this axial multi-intensity alternating projection method, the distance between the sample plane and the sensor plane is usually kept as close as possible, so that numerical propagation is used for projection instead of Fourier transform. Meanwhile, with a fixed position of the sensor, multiple intensity maps can also be recorded by radially moving the aperture near the sample, and then the complex field is recovered iteratively like the ER and HIO algorithms (Fig. 9b), the so-called ptychographic iterative engine (PIE)[17,18]. In this radial multi-intensity alternating projection method, each adjoining aperture constraint overlaps one another. Furthermore, angular multi-intensity alternating projection is also possible. By switching the aperture constraint from the spatial domain to the frequency domain, multiple intensity maps with different frequency information are recorded by changing the angle of the incident light (Fig. 9c), the so-called Fourier ptychography (FP)[19].

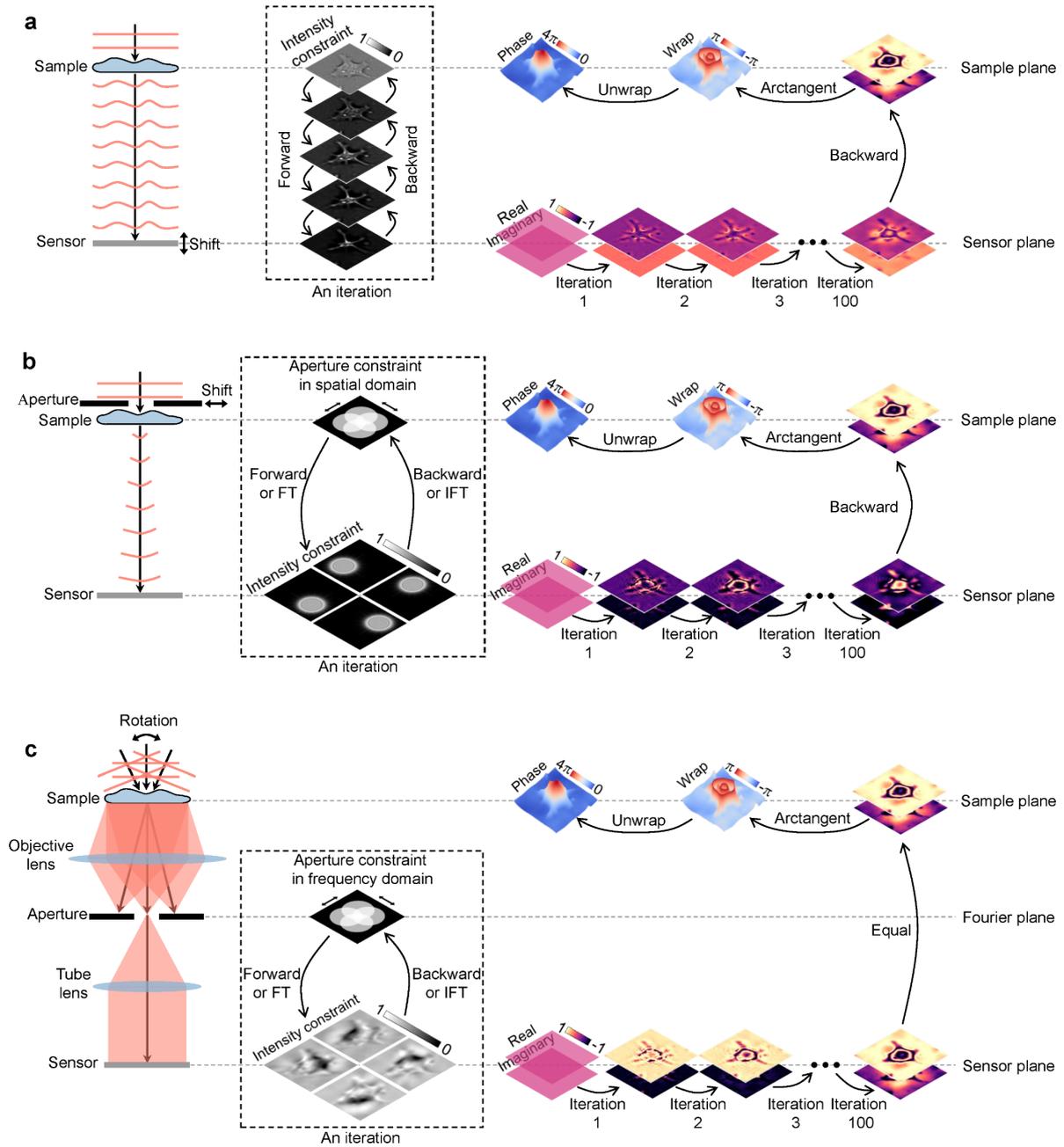

**Fig. 9 Description of multi-intensity alternating projection. a** Axial multi-intensity alternating projection. **b** Radial multi-intensity alternating projection. **c** Angular multi-intensity alternating projection. "Forward": forward numerical propagation. "Backward": backward numerical propagation.

In addition to alternating projections, there are two most representative non-convex optimization methods, namely the Wirtinger flow[36] and truncated amplitude flow algorithms[37]. They can be transformed into convex optimization problems through semidefinite programming, such as the PhaseLift algorithm[38].

**Deep learning (DL) for phase recovery.** In recent years, as an important step towards true artificial intelligence (AI), deep learning[39] has achieved unprecedented performance in many tasks of computer vision with the support of graphics processing units (GPUs) and large datasets. Similarly, since it was first used to solve the inverse problem in imaging in 2016[40], deep learning has demonstrated good potential in the field of computational imaging[41]. In the meantime, there is a rapidly growing interest in using deep learning for phase recovery (Fig. 10).

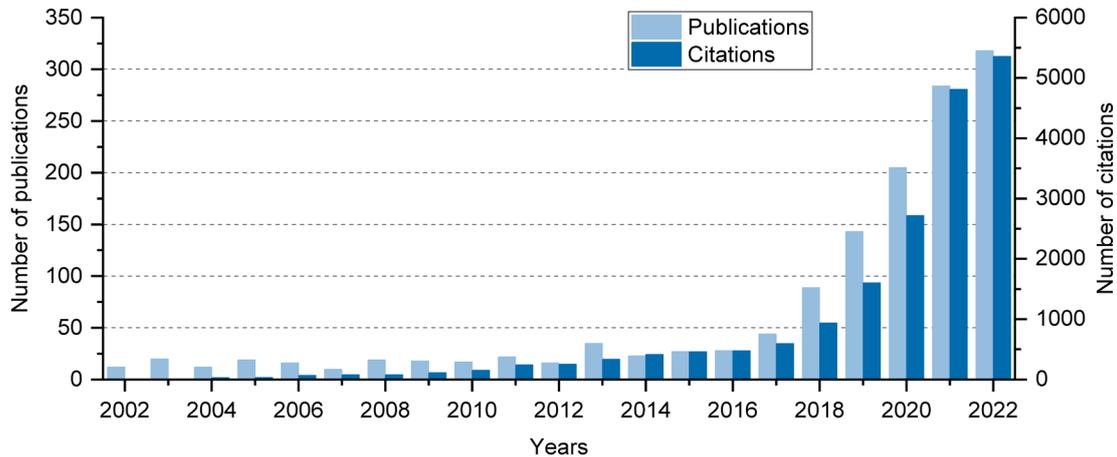

**Fig. 10 Growth in interest in using "deep learning for phase recovery" overtime is depicted by the number of publications and citations on Web of Science.** The used search code is "*TS=(("phase recovery" OR "phase retrieval" OR "phase imaging" OR "holography" OR "phase unwrapping" OR "holographic reconstruction" OR "hologram" OR "fringe pattern") AND ("deep learning" OR "network" OR "deep-learning"))*".

For the vast majority of "DL for PR", the implementation of deep learning is based on the training and inference of artificial neural networks (ANNs)[42] through input-label paired dataset, known as supervised learning (Fig. 11). In view of its natural advantages in image processing, the convolutional neural network (CNN)[43] is the most widely used ANN for phase recovery. Specifically, in order for the neural network to learn the mapping from physical quantity *A* to *B*, a large number of paired examples need to be collected to form a training dataset that implicitly contains this mapping relationship (Fig. 11a). Then, the gradient of the loss function is propagated backward through the neural network, and the network parameters are updated iteratively, thus internalizing this mapping relationship (Fig. 11b). After training, the neural network is used to compute $B_x$ from an unseen $A_x$ (Fig. 11c). In this way, deep learning has been used in all stages of phase recovery and phase processing.

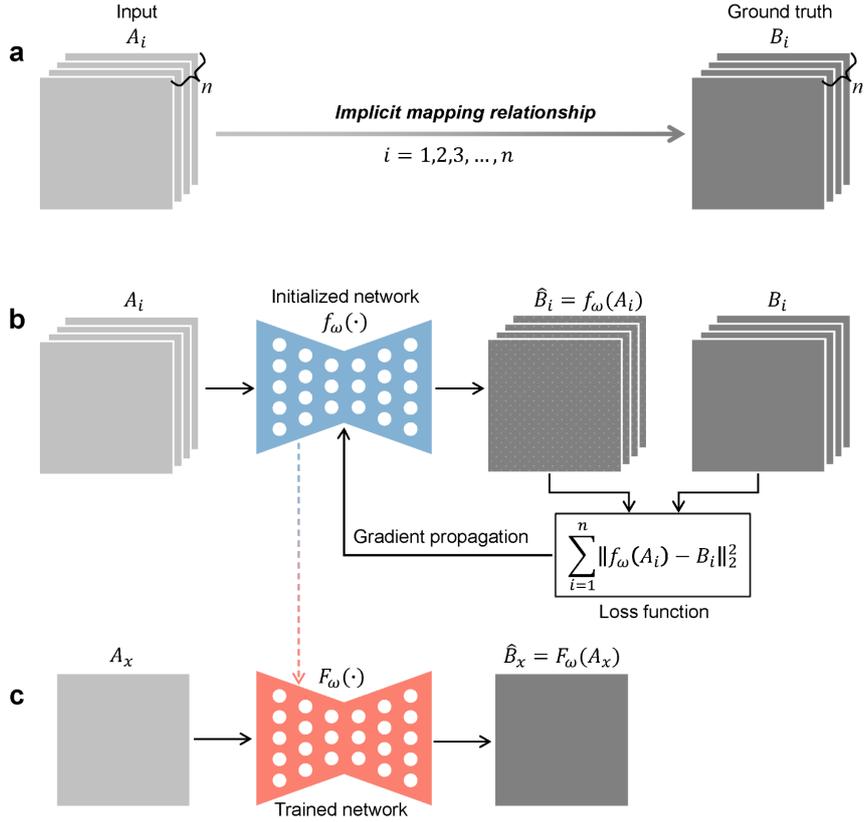

**Fig. 11 Implementation of deep learning with paired dataset and supervised learning. a** Datasets collection. **b** Network training. **c** Inference via a trained network. $\omega$: the parameters of the neural network, $n$: the sample number of the training dataset.

In fact, the rapid pace of deep-learning-based phase recovery has been documented in several excellent review papers. For example, Barbastathis *et al.*[44] and Rivenson *et al.*[45] reviewed how supervised deep learning powers the process of phase retrieval and holographic reconstruction. Zeng *et al.*[46] and Situ *et al.*[47] mainly focused on the use of deep learning in digital holography and its applications. Wang *et al.*[48] reviewed and compared different usage strategies of deep learning in phase unwrapping. Dong *et al.*[49] introduced a unifying framework for various algorithms and applications from the perspective of phase retrieval and presented its advances in machine learning. Differently, depending on where the neural network is used, we review various methods from the following four perspectives:

- In *DL-pre-processing for PR* (Section 2), the neural network performs some pre-processing on the intensity measurement before phase recovery, such as pixel super-resolution (Fig. 12a), noise reduction, hologram generation, and autofocusing.

- In *DL-in-processing for PR* (Section 3), the neural network directly performs phase recovery (Fig. 12b) or participates in the process of phase recovery together with the physical model or physics-based algorithm.
- In *DL-post-processing for PR* (Section 4), the neural network performs some post-processing after phase recovery, such as noise reduction (Fig. 12c), resolution enhancement, aberration correction, and phase unwrapping.
- In *DL for phase processing* (Section 5), the neural network uses the recovered phase for specific applications, such as segmentation (Fig. 12d), classification, and imaging modal transformation.

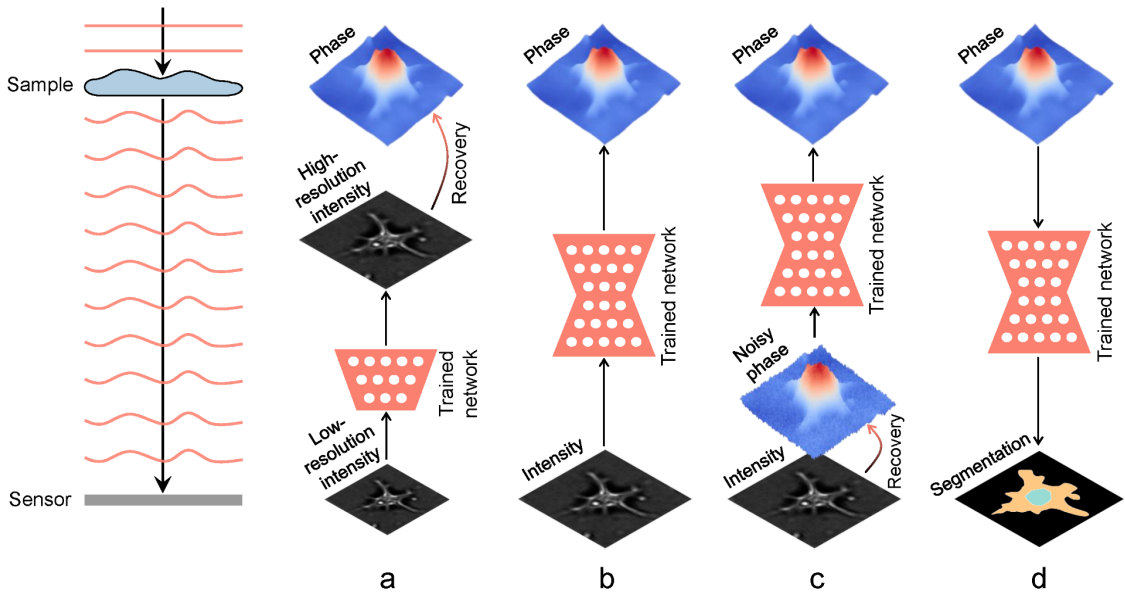

**Fig. 12 Overview example of "deep learning (DL) for phase recovery (PR) and phase processing". a** DL-pre-processing for PR. **b** DL-in-processing for PR. **c** DL-post-processing for PR. **d** DL for phase processing.

Finally, we summarize how to effectively use deep learning in phase recovery and look forward to potential development directions (Section 6). To let readers learn more about phase recovery, we present a live-updating resource (https://github.com/kqwang/phase-recovery).

## 2. DL-pre-processing for phase recovery

A summary of "DL-pre-processing for phase recovery" is presented in Table 1 and is described below, including pixel super-resolution (Section 2.1), noise reduction (Section 2.2), hologram generation (Section 2.3), and autofocusing (Section 2.4).

**Table 1 Summary of "DL-pre-processing for phase recovery"**

| Task | Reference | Input | Output | Network | Training dataset | Loss function |
|---|---|---|---|---|---|---|
| Pixel Super-resolution | Luo et al.[50] | Sub-pixel LR holograms | HR hologram | U-Net | Expt. and Sim.: 1,600 pairs | SSIM |
| | Byeon et al.[51] | LR hologram | HR hologram | SRCNN | Sim.: 192 pairs | $l_2$-norm |
| | Xin et al.[52] | LR hologram | HR hologram | Fast SRCNN | Sim.: 5,000 pairs | $l_2$-norm |
| | Ren et al.[53] | LR hologram | HR hologram | ResNet and SubPixelNet | Expt.: 800 pairs | $l_2$-norm |
| Noise reduction | Yan et al.[54] | Noisy fringe pattern | Noise-free fringe pattern | DnCNN | Sim.: 80,000 pairs | $l_1$-norm |
| | Lin et al.[55] | Noisy fringe pattern | Noise-free fringe pattern | CNN | Sim.: 230,400 pairs | $l_2$-norm |
| | Hao et al.[56] | Noisy fringe sub-pattern | Noise-free fringe sub-pattern | FFDNet | Sim.: 1,200 pairs | $l_2$-norm |
| | Zhou et al.[57,58] | Noisy fringe pattern | Noise-free fringe pattern | Spectral CNN | Sim.: 1,200 pairs | --- |
| | Reyes-Figueroa et al.[59] | Noisy fringe pattern | Noise-free fringe pattern | U-Net and ResNet | Sim.: 25,000 pairs | $l_1$-norm |
| | Gurrola-Ramos et al.[60] | Noisy fringe pattern | Noise-free fringe pattern | U-Net and DenseNet | Sim.: 1,500 pairs | $l_1$-norm |
| Hologram generation | Zhang et al.[61,62] | Hologram | Phase-shifting holograms | Y-Net | Sim.: --- | $l_2$-norm |
| | Yan et al.[63] | Hologram | Single phase-shifting hologram | ResNet | Sim: 12,000 pairs | GAN loss |
| | Zhao et al.[64] | Hologram | Phase-shifting holograms | MPRNet | Sim.: --- | Charbonnier and Edge |
| | Huang et al.[65] | Hologram | Phase-shifting holograms | Y-Net | Expt.:4,000 pairs | $l_2$-norm |
| | Wu et al.[66] | Hologram | Single phase-shifting hologram | U-Net | Sim.: 6,400 pairs | GAN loss |
| | Luo et al.[67] | Hologram | Multi-distance holograms | U-Net | Sim.: 440 pairs | GAN loss |
| | Li et al.[68] | Hologram | Hologram with another wavelength | U-Net | Sim.: 20,000 pairs | GAN loss |
| | Li et al.[69] | Two holograms | Hologram with another wavelength | Y-Net | Sim.: 8,000 pairs | $l_1$-norm |
| | Xu et al.[70] | dual-wavelength hologram | Two single-wavelength holograms | U-Net | Sim.: 1,800 pairs | $l_2$-norm |
| Autofocusing | Pitkäaho et al.[71] | Hologram | Defocus distance (21 types) | AlexNet | Expt. and sim.: 485,856 pairs | Cross entropy |
| | Ren et al.[72] | Hologram | Defocus distance (5 types) | CNN | Expt.: > 5,000 pairs | Cross entropy |
| | Son et al.[73] | Hologram | Defocus distance (10 types) | CNN | Sim.: 40,180 pairs | Cross entropy |
| | Couturier et al.[74] | Hologram | Defocus distance (101 types) | DenseNet | Expt.: 7,000 pairs | Cross entropy |
| | Ren et al.[75] | Hologram (amplitude or phase object) | Defocus distance | CNN | Expt.: 5,000 and 2,000 pairs | $l_1$-norm |
| | Pitkäaho et al.[76] | Hologram (cells) | Defocus distance | AlexNet and VGG | Expt.: 437,271 pairs | $l_2$-norm |
| | Jaferzadeh et al.[77] and Moon et al.[78] | Hologram (single cell) | Defocus distance | CNN | Expt.: 3,000 and 2,400 pairs | $l_2$-norm |
| | Tang et al.[79] | Fixed tensor | Defocus distance | MLP (untrained) | Expt.: 1 | $l_2$-norm |
| | Cuenat et al.[80,81] | Hologram (USAF 1951) | Defocus distance | ViT | Expt.: 104,400 pairs Sim.: 40,000 pairs | log cosh |
| | Lee et al.[82] | Spatial spectrum | Defocus distance | CNN | Sim.: --- | $l_2$-norm |
| | Shimobaba et al.[83] | 1/4 power spectrum | Defocus distance | CNN | Sim.: --- | $l_2$-norm |

"---" indicates not available. "LR" is short for low-resolution. "HR" is short for high-resolution. "Expt." is short for experiment. "Sim." is short for simulation. "GAN loss" means training the network in an adversarial generative way. "MLP" is short for multi-layer perceptron.

## 2.1 Pixel super-resolution

A high-resolution image generally reveals more detailed information about the object of interest. Therefore, it is desirable to recover a high-resolution image from one or multiple low-resolution measurements of the same field of view, a process known as pixel super-resolution. Similarly, from multiple sub-pixel-shifted low-resolution holograms, a high-resolution hologram can be recovered by pixel super-resolution algorithms[84]. Luo et al.[50] proposed to use the U-Net for this purpose. Compared with iterative pixel super-resolution algorithms, this deep learning method has an advantage in inference time while ensuring a same level of resolution improvement, and maintains high performance even with a reduced number of input low-resolution holograms.

After the pixel super-resolution CNN (SRCNN) was proposed for single-image super-resolution in the field of image processing[85], this type of deep learning method was also used in other optical super-resolution problems, such as brightfield microscopy[86] and fluorescence microscopy[87]. Similarly, this method of inferring corresponding high-resolution images from low-resolution versions via deep neural networks can also be used for holograms pixel super-resolution before doing phase recovery by conventional recovery methods (Fig. 13).

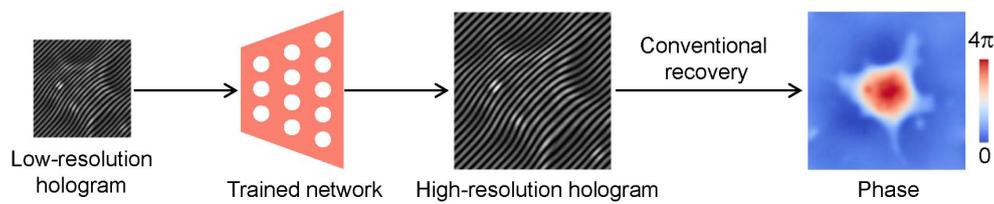

Fig. 13 Description of deep-learning-based hologram super-resolution.

Byeon et al.[51] first applied the SRCNN to hologram pixel super-resolution, and named it HG-SRCNN. Compared with conventional focused-image-trained SRCNN and bicubic interpolation, this method, trained with defocus in-line holograms, can infer higher-quality high-resolution holograms. Xin et al.[52] used an improved fast SRCNN (FSRCNN) to do pixel super-resolution for white-light holograms, significantly improving the identification and accuracy of three-dimensional (3D) measurement results. Under the premise of improved accuracy, the inference speed of FSRCNN is nearly ten times faster than that of SRCNN.

Ren et al.[53] proposed to use a CNN, incorporating the residual network (ResNet) and sub-pixel network (SubPixelNet), for pixel super-resolution of a single off-axis hologram. They found that compared to $l_1$-norm and structural similarity index (SSIM)[88], the neural network trained using $l_2$-norm as the loss function performed best. Moreover, this deep learning method reconstructs high-resolution off-axis holograms with better quality, compared with conventional image super-resolution methods, such as bicubic, bilinear, and nearest-neighbor interpolations.

## 2.2 Noise reduction

Most phase recovery methods, especially holography, are performed with a coherent light source; therefore, coherent noise is an unavoidable issue. In addition, noise can be caused by environmental disturbances and the recording process of the image sensor. Therefore, it is very important to reduce the noise from the hologram before phase recovery. Filter-based methods, such as windowed Fourier transform (WFT)[89], have been widely used in hologram noise reduction, but most of these methods face a trade-off between good filtering performance and time cost.

In 2017, Zhang et al.[90] opened the door to image denoising using the deep CNN, called DnCNN. Subsequently, the DCNN was introduced to the field of fringe analysis for fringe pattern denoising (Fig. 14).

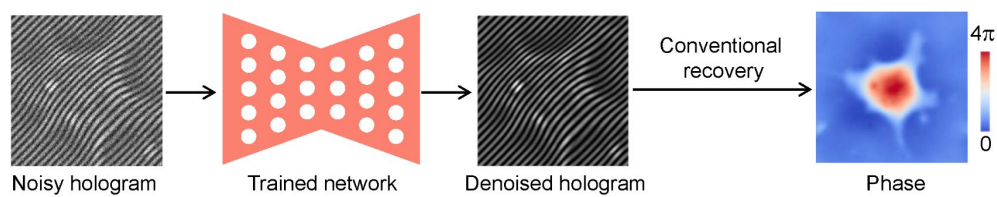

Fig. 14 Description of deep-learning-based hologram noise reduction.

Yan et al.[54] first applied the DnCNN to fringe pattern denoising, which has higher precision around image boundaries and needs less inference time than WFT. Similar conclusions can also be seen in the work of Lin et al.[55]. Then, inspired by the FFDNet[91], Hao et al.[56] downsampled the input fringe pattern into four sub-images before using the DnCNN for denoising, leading to a faster inference speed. Furthermore, Zhou et al.[57,58] converted this batch-denoising DnCNN into the frequency domain. Specifically, they first computed the Fourier transform of the downsampled sub-images, and then used the DnCNN to achieve noise reduction in the frequency domain, and finally applied upsampling and inverse Fourier transform to obtain the denoised fringe pattern. From the comparison results, their method outperforms that of Yan et al. and Hao et al. at different noise levels. Reyes-Figueroa et al.[59] further showed that the U-Net and its improved version (V-Net) are better than DnCNN for fringe pattern denoising, because their proposed V-Net has more channels on the outer side than on the inner side, retaining more details. Given the U-Net's outstanding mapping capabilities, Gurrola-Ramos et al.[60] also improved it for fringe pattern denoising, where dense blocks are leveraged for reusing feature layers, local residual learning is used to address the vanishing gradient problem, and global residual learning is used to estimate the noise of the image instead of the denoised image directly. Compared with other neural networks mentioned above, it has a minor model complexity while maintaining the highest accuracy.

2.3 Hologram generation

As mentioned in Introduction, in order to recover the phase, multiple intensity maps are needed in many cases, such as phase-shifting holography and axial multi-intensity alternating projection. Given its excellent mapping capability, the neural network can be used to generate other relevant holograms from known ones, thus enabling phase recovery that requires multiple holograms (Fig. 15). In this approach, the input and output usually belong to the same imaging modality with high feature similarity, so it is easier for the neural network to learn. Moreover, the dataset is collected only by experimental record or simulation generation, without the need for phase recovery as ground-truth in advance by conventional methods.

Zhang *et al.*[61,62] first proposed the idea of generating holograms with holograms before phase recovery with the conventional method (Fig. 15a). From a single hologram, the other three holograms with $\pi/2$, $\pi$, and $3\pi/2$ phase shifts were simultaneously generated by the Y-Net[92], and then phase recovery was implemented by the four-step phase-shifting method. The motivation to infer holograms instead of phase via a network is that for different types of samples, the spatial differences between their holograms were significantly lower than that of their phase. Accordingly, this phase recovery based on the hologram generation has better generalization ability than recovering phase from holograms directly with the neural network, especially when the spatial characteristics differences of the phase between the training and testing datasets are relatively large[62]. Since the phase-shift between the generated holograms are equal, Yan *et al.*[63] proposed to generate noise-free phase-shifting holograms using a simple end-to-end generative adversarial network (GAN) in a manner of sequential concatenation. Subsequently, for better performance in balancing spatial details and high-level semantic information, Zhao *et al.*[64] applied the multi-stage progressive image restoration network (MPRNet)[93] for phase-shifting hologram generation. Huang *et al.*[65] and Wu *et al.*[66] then expanded this approach from four-step to three-step and two-step phase-shifting methods, respectively.

Luo *et al.*[67] proposed to generate holograms with different defocus distances from one hologram via a neural network, and then achieve phase recovery with alternating projection (Fig. 15b). Similar to the work of Zhang *et al.*[62], they proved that the use of neural networks with less difference between the source domain and the target domain could enhance the generalization ability. As for multi-wavelength holography, Li *et al.*[68,69] harnessed a neural network to generate a hologram of another wavelength from one or two holograms of known wavelength, thereby realizing two-wavelength and three-wavelength holography. At the same time, Xu *et al.*[70] realized a one-shot two-wavelength and three-wavelength holography by generating the corresponding single-wavelength holograms from a two-wavelength or three-wavelength hologram with information crosstalk.

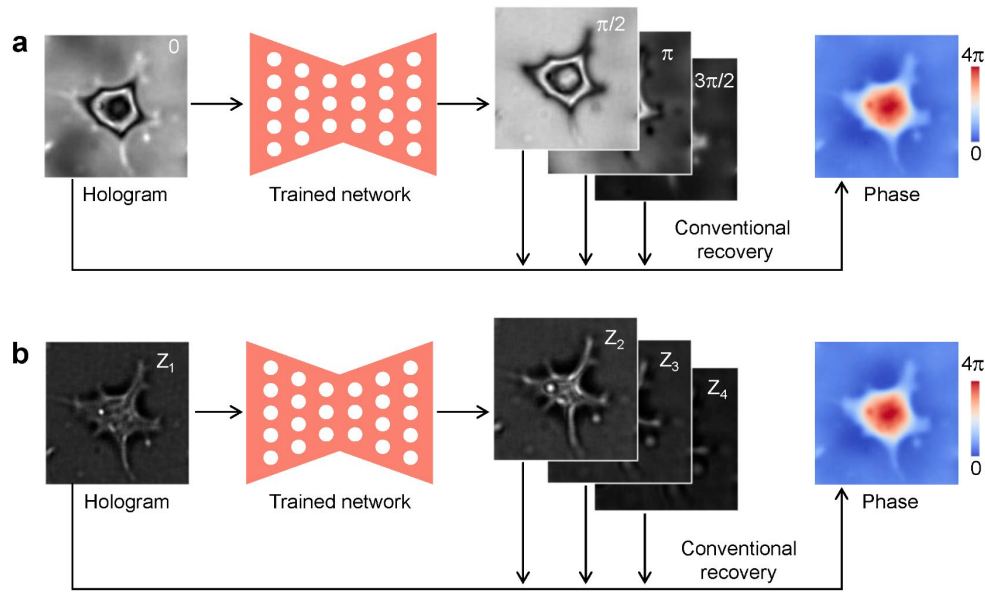

Fig. 15 Description of deep-learning-based hologram generation for (a) phase-shifting method and (b) axial multi-intensity alternating projection method.

2.4 Autofocusing

In lensless holography, the phase of the sample plane can only be recovered if the distance between the sensor plane and the sample plane is known. Defocus distance estimation thus becomes a fundamental problem in holography, which is also known as autofocusing.

Deep learning methods for autofocus essentially use the neural network to estimate the defocus distance from the hologram (Fig. 16), which can be regarded as either a classification problem[71–74] or a regression problem[75–78,80–83].

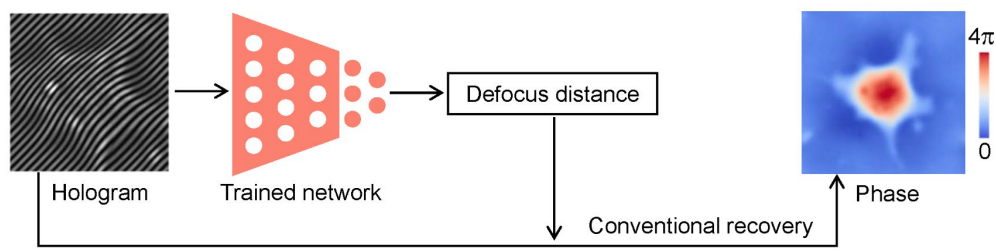

Fig. 16 Description of deep-learning-based hologram numerical refocusing.

From the perspective of classification, Pitkäaho et al.[71] first proposed to estimate the defocus distance from the hologram by a CNN. In their scheme, the zero-order and twin-image terms need to be removed before the trained neural network classifies the holograms into different discrete defocus distances. Meanwhile, Ren et al.[72] advocate directly using raw holograms collected at different defocus distances as the input of the neural networks. Furthermore, they revealed the advantages of neural networks over other machine learning algorithms in the task of autofocusing. Immediately afterward, Son et al.[73] also verified the feasibility of autofocus by classification through numerical simulations. Subsequently, Couturier et al.[74] improved the accuracy of defocus distance estimation by using a deeper CNN for categorizing defocus distance into a greater number of classes.

Nevertheless, no matter how many classes there are, the defocus distance estimated by these classification-based methods is also discrete, which is still not precise enough in practice. Thus, Ren et al.[75] further developed an approach to treat the defocus distance estimation as a regression problem, where the output of the neural network is continuous. They verified the superiority of this deep-learning-based regression method with amplitude samples and phase samples, respectively, and tested the adaptability under different exposure times and incident angles. Later, Pitkäaho et al.[76] also extended their previous classification-based work[71] to this regression-based approach. While these methods estimate the defocus distance of the entire hologram, Jaferzadeh et al.[77] and Moon et al.[78] proposed to take out the region of interest from the whole hologram as the input to estimate the defocus distance. In order to get rid of the constraint of known defocus distance as the label of the training dataset, Tang et al.[79] proposed to iteratively infer the defocus distance by an untrained network with a defocus hologram and its in-focus phase. Later on, Cuenat et al.[81] demonstrated the superiority of the vision Transformer (ViT) over typical CNNs in defocus distance estimation. Because the spatial spectrum information is also helpful for the defocus distance estimation[94], Lee et al.[82] and Shimobaba et al.[83] proposed to use the spatial spectrum or power spectrum of holograms as the network input to estimate the defocus distance.

## 3. DL-in-processing for phase recovery

In "DL-in-processing for phase recovery", the neural network directly performs the inference process from the measured intensity image to the phase (network-only strategy in Section 3.1), or together with the physical model or physics-based algorithm to achieve the inference (network-with-physics strategy in Section 3.2).

## 3.1 Network-only strategy

The network-only strategy uses a neural network to perform phase recovery, where the network input is the measured intensity image and the output is the phase. A summary of various methods is presented in Table 2 and described below, where we classify them into dataset-driven (DD) approaches and physics-driven (PD) approaches.

**Table 2 Summary of network-only strategy**

| Task | Reference | Input | Output | Network | Training dataset | Loss function |
|---|---|---|---|---|---|---|
| Dataset-driven (DD) approach | Sinha et al.[95] | Diffraction image | Phase | U-Net and ResNet | Expt.: 10,000 pairs | $l_1$-norm |
| | Li et al.[96] | Diffraction image | Phase | U-Net and ResNet | Expt.: 10,000 pairs | NPCC |
| | Deng et al.[97] | Diffraction image | Phase | U-Net and ResNet | Expt.: 10,000 pairs | NPCC |
| | Goy et al.[98] | Weak-light diffraction | Phase | U-Net and ResNet | Expt.: 9,500 pairs | NPCC |
| | Wang et al.[99] | In-line hologram | Phase | U-Net and ResNet | Expt.: 9000 and 11623 pairs | $l_2$-norm |
| | Nguyen et al.[100] | Multiple LR intensity images (FP) | HR phase | U-Net and DenseNet | Expt.: --- | GAN loss and $l_1$-norm |
| | Cheng et al.[101] | LR intensity image (FP) | HR phase and amplitude | CNN and ResNet | Expt.: 20 fields-of-view | $l_2$-norm |
| | Cherukara et al.[102] | Far-field diffraction | Phase or amplitude | SegNet (two) | Sim.: 180,000 pairs | Cross-entropy |
| | Ren et al.[103] | Off-axis hologram | Phase or amplitude | ResNet and SubPixelNet | Expt.: >10,000 pairs | $l_2$-norm |
| | Yin et al.[104] | Hologram | Phase | U-Net | Expt.: 2,400 and 200-2,000 (unpaired) | Cycle-GAN loss |
| | Lee et al.[105] | Hologram | Phase and amplitude | U-Net and CNN | Expt.: 6000-9,060 (unpaired) | Cycle-GAN loss and SSIM |
| | Hu et al.[106] | Spots' intensity image | Phase | U-Net and ResNet | Sim.: 46,080 pairs | $l_2$-norm |
| | Wang et al.[107] | Defocus intensity image | Phase | U-Net and ResNet | Expt.: 20,037 pairs | $l_2$-norm |
| | Pirone et al.[108] | Hologram in different angles | Phase | CAN | Expt.: 4,000 pairs | $l_1$-norm |
| | Xue et al.[109] | Bright- and dark-field images | Phase | U-Net and BNN | Expt.: 185 groups | $l_1$-norm and uncertainty term |
| | Li et al.[110] | Two images of symmetric illumination | Phase | U-Net | Sim.: 1,301 groups | GAN loss |
| | Wang et al.[92,111] | Hologram | Phase and amplitude | Y-Net | Expt.: 1,331 pairs | $l_2$-norm |
| | Zeng et al.[112] | Hologram | Phase or amplitude | CapsNet | Expt.: --- | $l_2$-norm |
| | Wu et al.[113] | Far-field diffraction | Phase and amplitude | Y-Net | Sim.: 142,500 groups | Loss in real and reciprocal space |
| | Huang et al.[114] | Two or 3 holograms | Complex field | U-Net and Recurrent CNN | Expt.: 208 groups | GAN loss and $l_1$-norm and SSIM |
| | Uelwer et al.[115] | Far-field diffraction | Phase | Cascaded neural network | Sim.: --- | $l_2$-norm or $l_1$-norm |
| | Castaneda et al.[116] | Off-axis hologram | Wrapped phase | U-Net | Expt.: 1,512 pairs | GAN loss and TSM and STD |
| | Jaferzadeh et al.[117] | Off-axis hologram | Phase | U-Net | Expt.: 900 pairs | GAN loss |
| | Luo et al.[118] | Hologram | Phase | MCN | Expt.: 1 pair | Bucket error rate (BER) loss |
| | Ding et al.[119] | LR image | HR phase | U-Net and ViT | Expt.: 3,500 and 3,500 (unpaired) | Cycle-GAN loss |
| | Ye et al.[120] | Far-field diffraction | Complex field | MLP and CNN | Sim. and Expt.: --- | $l_1$-norm |
| | Chen et al.[121,122] | Three or 4 holograms | Complex field | ResNet and Fourier module (FIN) | Expt.: 600 groups | $l_1$-norm, complex domain and perceptual loss |

| | | | | | |
|---|---|---|---|---|---|
| | Shu et al.[123] | Hologram | Phase | Network based on NAS | Expt.: 276 pairs | MixGE and binary and sparsity loss |
| Physics-driven (PD) approach | Boominathan et al.[124] | LR intensity images (FP) | HR Phase and amplitude | U-Net | Sim.: 1 (input only) | $l_2$-norm with physical model |
| | Wang et al.[125] | Diffraction image | Phase | U-Net | Sim. and Expt.: 1 (input only) | $l_2$-norm with physical model |
| | Zhang et al.[126] | Diffraction image | Phase | U-Net | Sim. and Expt.: 1 (input only) | $l_2$-norm with defocus distance and physical model |
| | Yang et al.[127,128] | Diffraction image | Phase and amplitude | U-Net | Sim. and Expt.: 1-180 (input only) | $l_2$-norm with aperture constraint |
| | Bai et al.[129] | Hologram | dual-wavelength Phase | CDD | Expt.: 1 (input only) | $l_2$-norm with physical model |
| | Galande et al.[130] | Hologram | Phase and amplitude | U-Net | Expt.: 1 (input only) | $l_2$-norm with physical model and denoiser |
| | Yao et al.[131] | 3D diffraction image | Phase and amplitude | 3D Y-Net | Sim.: 52,000 (input only) | $l_2$-norm with physical model |
| | Li et al.[132] | Two diffraction images | Phase | Two-to-one Y-Net | Sim.: 500 (input only) | $l_2$-norm with physical model |
| | Bouchama et al.[133] | LR intensity images (FP) | HR Phase and amplitude | U-Net | Sim.: 10,000 (input only) | $l_2$-norm with physical model |
| | Huang et al.[134] | Two holograms | Phase and amplitude | --- | Sim.: 100,000 (input only) | $l_2$-norm and Fourier domain $l_1$-norm |

**Dataset-driven approach.** As one of the most commonly adopted strategies, data-driven deep learning phase recovery methods presuppose a large number of paired input-label datasets. Usually, it is necessary to experimentally collect a significant number of intensity images (such as diffraction images or holograms, etc.) as input, and use conventional methods to calculate the corresponding phase as ground-truth (Fig. 17a). The key lies in that this paired dataset implicitly contains the mapping relationship from intensity to phase. Then, an untrained/initialized neural network is iteratively trained with the paired dataset as an *implicit prior*, where the gradient of the loss function propagates into the neural network to update the parameters (Fig. 17b). After training, the network is used as an end-to-end mapping to infer the phase from intensity (Fig. 17c). Therefore, the DD approach is to guide/drive the training of the neural network with this implicit mapping, which is internalized into the neural network as the parameters are iteratively updated.

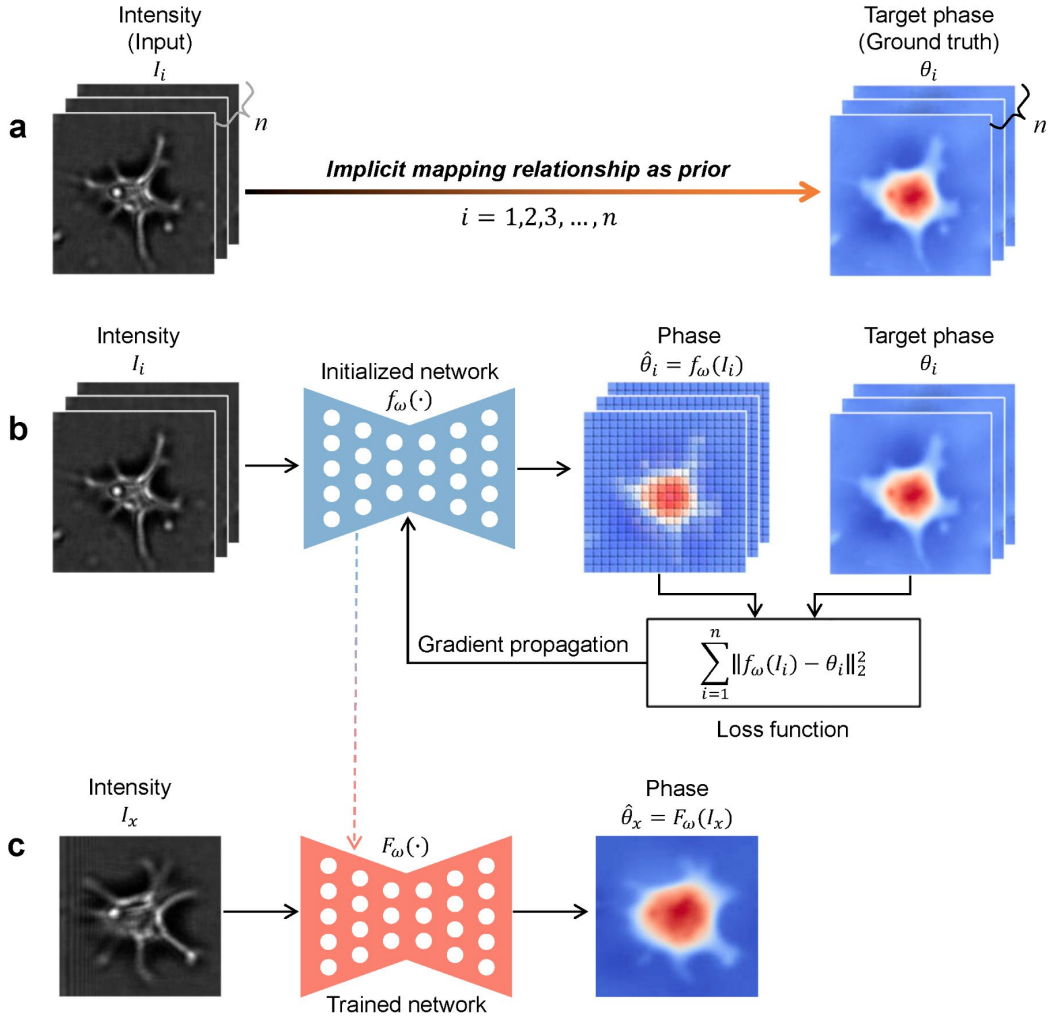

**Fig. 17 Description of dataset-driven approach for phase recovery. a** Dataset collection. **b** Network training. **c** Inference via a trained network.

Sinha *et al.*[95] were among the first to demonstrate this end-to-end deep learning strategy for phase recovery, in which the phase of objects is inferred from corresponding diffraction images via a trained deep neural network. In dataset collection, they used a phase-only spatial light modulator (SLM) to load different public image datasets to generate the phase as ground-truth, and after a certain distance, place the image sensor to record the diffraction image as input. The advantage is that both the diffraction image and the phase are known, and is easily collected in large quantities. Through comparative tests, they verified the adaptability of the deep neural network to unseen types of datasets and different defocus distances. Although this scheme cannot be used in practical application due to the use of the phase-type spatial light modulator, their pioneering work opens the door to deep-learning-inference phase recovery. For instance, Li *et al.*[96] introduced the negative Pearson correlation coefficient (NPCC)[135] as a loss function to train the neural network, and enhanced the spatial resolution by a factor of two by flattening the power spectral density of the training dataset. Deng et al.[97] found that the higher the Shannon entropy of the training dataset, the stronger the generalization ability of the trained neural network. Goy et al.[98] extended the work to phase recovery under weak-light illumination.

Meanwhile, Wang et al.[99] extended the diffraction device of Sinha et al.[95] to an in-line holographic device by adding a coaxial reference beam, and used the in-line hologram instead of the diffraction image as the input to a neural network for phase recovery. Nguyen et al.[100] applied this end-to-end strategy for FP, inferring the high-resolution phase from a series of low-resolution intensity images via a U-Net, and Cheng et al.[101] further used a single low-resolution intensity image under optimized illumination as the neural network input. Cherukara et al.[102] extended this end-to-end deep learning strategy to CDI, in which they trained two neural networks with simulation datasets to infer the amplitude or phase of objects from far-field diffraction intensity maps, respectively. Ren et al.[103] demonstrated the time and accuracy superiority of this end-to-end deep learning strategy over conventional numerical algorithms in the case of off-axis holography. Yin et al.[104] introduced the cycle-GAN to extend this end-to-end deep learning strategy to the application scenario of unpaired datasets. Lee et al.[105] replaced the forward generator of the cycle-GAN by numerical propagation, improving the phase recovery robustness of neural networks in highly perturbative configurations. Hu et al.[106] applied this end-to-end deep learning strategy to the Shack-Hartmann wavefront sensor, inferring the phase directly from a spot intensity image after the micro-lens array. Wang et al.[107] extended this end-to-end deep learning strategy to TIE, using a trained neural network to infer the phase of the cell object from a defocus intensity image illuminated by partially coherent light. Pirone et al.[108] applied this hologram-to-phase deep learning strategy to improve the reconstruction speed of 3D optical diffraction tomography (ODT) from tens of minutes to a few seconds. Tayal et al.[136] demonstrated the use of data augmentation and a symmetric invariant loss function to break the symmetry in the end-to-end deep learning phase recovery.

In addition to expanding the application scenarios of this end-to-end deep learning strategy, some researchers focused on the performance and advantages of different neural networks in phase recovery. Xue et al.[109] applied Bayesian neural network (BNN) into FP for inferring model uncertainty while doing phase recovery. Li et al.[110] applied GAN for phase recovery, inferring the phase from two symmetric-illumination intensity images. Wang et al.[92,111] proposed a one-to-multi CNN, Y-Net[92], from which the amplitude and phase of an object can be inferred from the input intensity simultaneously. Zeng et al.[112] introduce the capsule network to overcome information loss in the pooling operation and internal data representation of CNN. Compared with conventional CNN, their proposed capsule-based CNN (RedCap) saves 75% of network parameters while ensuring higher holographic reconstruction accuracy. Wu et al.[113] applied the Y-Net[92] to CDI for simultaneous inference of phase and amplitude. Huang et al.[114] introduced a recurrent convolution module into U-Net, trained using GAN, for holographic reconstruction with autofocus. Uelwer et al.[115] used a cascaded neural network for end-to-end phase recovery. Castaneda et al.[116] and Jaferzadeh et al.[117] introduced GAN into off-axis holographic reconstruction. Luo et al.[118] added dilated convolutions into a CNN, termed mixed-context network (MCN)[118], for phase recovery. By comparing in a one-sample-learning scheme, they found that MCN is more accurate and compact than the conventional U-Net. Ding et al.[119] added ViT into U-Net and trained it with low-resolution intensity as input and high-resolution phase as ground-truth using cycle-GAN. The trained neural network can do phase recovery while enhancing the resolution, and has higher accuracy than the conventional U-Net. In CDI, Ye et al.[120] used a multi-layer perceptron for feature extraction before a CNN, considering the property of the far-field (Fourier) intensity images where the data are globally correlated. Chen et al.[121,122] combined the spatial Fourier transform module with ResNet, termed Fourier imager network (FIN), to achieve holographic reconstruction with superior generalization to new types of samples and faster inference speed (9-fold faster than their previous recurrent neural network, 27-fold faster than conventional iterative algorithms). Shu et al.[123] applied neural architecture search (NAS) to automatically optimize the network architecture for phase recovery. Compared with the conventional U-Net, the peak signal-to-noise ratio (PSNR) of their NAS-based network is increased from 34.7 dB to 36.1 dB, and the inference speed is increased by 27-fold.

As a similar deep learning phase recovery strategy in adaptive optics, researchers demonstrated that neural networks could be used to infer the phase of the turbulence-induced aberration wavefront or its Zernike coefficient from the distortion intensity of target objects[137]. In these applications, only the wavefront that is subsequently used for aberration correction is of interest, not the RI distribution of turbulence itself that produces this aberration wavefront.

**Physics-driven approach.** Different from the dataset-driven approach that uses input-label paired dataset as an implicit prior for neural network training, physical models, such as numerical propagation, can be used as an *explicit prior* to guide/drive the inference or training of neural networks, termed physics-driven (PD) approach. On the one hand, this explicit prior can be used to iteratively optimize an untrained neural network to infer the corresponding phase and amplitude from the measured intensity image as input, referred to as the untrained PD (uPD) scheme (Fig. 18a). On the other hand, this explicit prior can be used to train an untrained neural network with a large number of intensity images as input, which then can infer the corresponding phase from unseen intensity images, an approach called the trained PD (tPD) scheme (Fig. 18b).

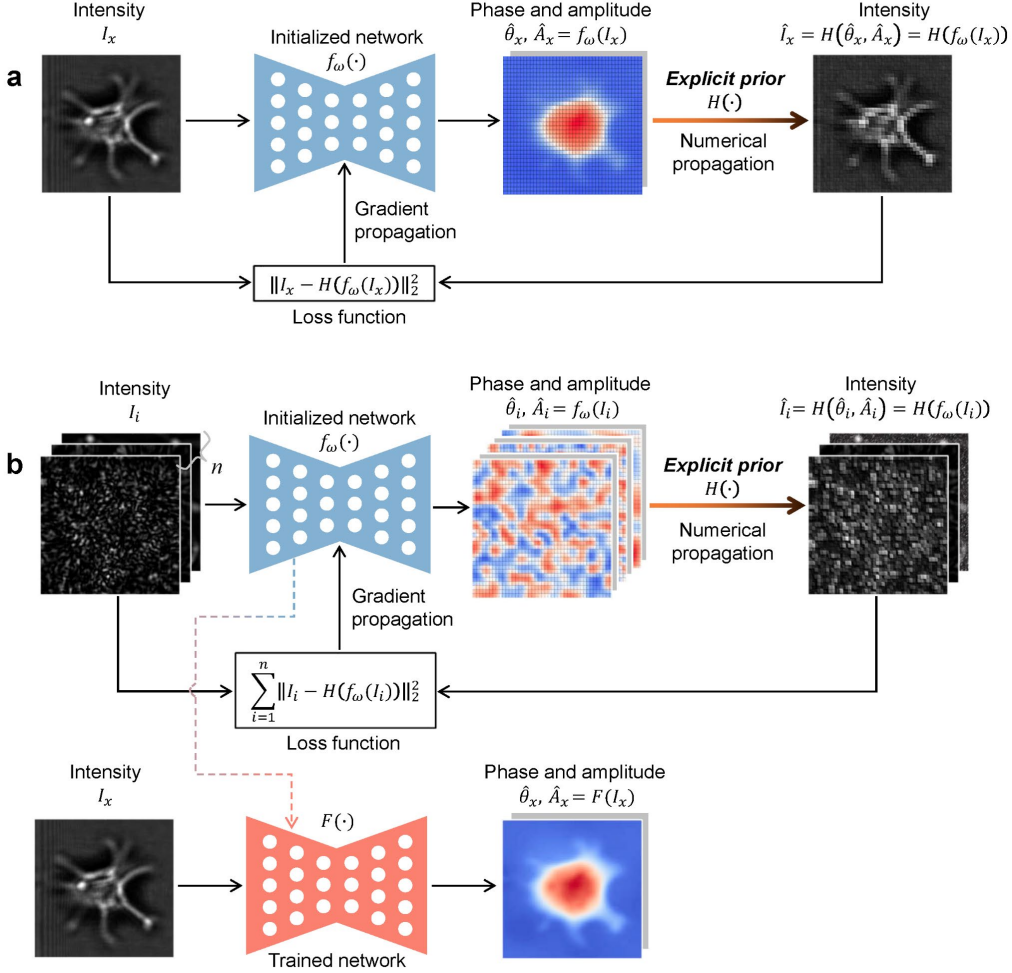

**Fig. 18 Description of physics-driven approach for phase recovery. a** Untrained PD (uPD) scheme. **b** Trained PD (tPD) scheme.

In order to more intuitively understand the difference and connection between the DD and PD approaches, let us compare the loss functions in Fig. 17 and Fig. 18:

$$Loss_{DD} = \sum_{i=1}^{n} \|f_\omega(I_i) - \theta_i\|_2^2, \tag{1}$$

$$Loss_{uPD} = \|I_x - H(f_\omega(I_x))\|_2^2, \tag{2}$$

$$Loss_{tPD} = \sum_{i=1}^{n} \|I_i - H(f_\omega(I_i))\|_2^2, \tag{3}$$

where $\|\cdot\|_2^2$ denotes the square of the $l_2$-norm (or other distance functions), $f_\omega(\cdot)$ is a neural network with trainable parameters $\omega$, $H(\cdot)$ is a physical model (such as numerical propagation, Fourier transform, or FP measurement model), $I_i$ is the measured intensity image in the training dataset, $\theta_i$ is the phase in the training dataset, $I_x$ is the measured intensity image of a test sample, and $n$ is the number of samples in the training dataset. In Eq. (1) for the DD approach, the priors used for network training are the measured intensity image and corresponding ground-truth phase. Meanwhile, in Eqs. (2) and (3) for the PD approaches, the priors used for network inference or training are the measured intensity image and physical model, instead of the phase.

This PD approach was first implemented in the work on Fourier Ptychography by Boominathan et al.[124]. They proposed it in the higher overlap case, including the scheme of directly using an untrained neural network for inference (uPD) and the scheme of training first and then inferring (tPD), and demonstrated the former by simulation.

For the uPD scheme, Wang et al.[125] used a U-Net-based scheme to iteratively infer the phase of an object from a measured diffraction image whose de-focus distance is known. Their method demonstrates higher accuracy than conventional algorithms (such as GS and TIE) and the DD scheme, at the expense of a longer inference time (about 10 minutes for an input with 256 × 256 pixels). Zhang et al.[126] extended this work to the case where the defocus distance is unknown, by including it as another unknown parameter together with the phase to the loss function. Yang et al.[127,128] further generalized this to the complex field inference by introducing an aperture constraint into the loss function, and pointed out that it would cost as much as 600 hours to infer 3,600 diffraction images with this uPD scheme. Meanwhile, Bai et al.[129] extended this from a single-wavelength case to a dual-wavelength case. Galande et al.[130] found that this way of neural network optimization with a single-measurement intensity input lacks information diversity and can easily lead to overfitting of the noise, which can be mitigated by introducing an explicit denoiser. This way of using the object-related intensity image as the neural network input makes it possible to internalize the mapping relationship between intensity and phase into the neural network through pre-training. It is worth mentioning that some researchers proposed to make adjustments to the uPD scheme, using the initial phase and amplitude recovered by backward numerical propagation as the neural network input[138–140], which reduces the burden on the neural network to obtain higher inference accuracy.

Although the phase can be inferred from the measured intensity image through an untrained neural network without any ground-truth, the uPD scheme inevitably requires a large number of iterations, which excludes its use in many dynamic applications. Therefore, to adapt the PD scheme to dynamic inference, Yang et al.[127,128] adjusted their previously proposed uPD scheme to the tPD scheme by pre-training the neural network using a small part of the measured diffraction images, and then using the pre-trained neural network to infer the remaining ones. Yao et al.[131] trained a 3D version of the Y-Net[92] with simulated diffraction images as input, and then used the pre-trained neural network for direct inference or iterative refinement, which is 100 and 10 times faster than conventional iterative algorithms, respectively. Li et al.[132] proposed a two-to-one neural network to reconstruct the complex field from two axially displaced diffraction images. They used 500 simulated diffraction images to pre-train the neural network, and then inferred an unseen diffraction image by refining the pre-trained neural network for 100 iterations. Bouchama et al.[133] further extended the tPD scheme to Fourier Ptychography of low overlap cases by simulated datasets. Different from the above ways of generating training datasets from natural images or real experiments, Huang et al.[134] proposed to generate training datasets from randomly synthesized artificial images with no connection or resemblance to real-world samples. They further trained a neural network with this generated dataset and a physics-consistency loss, which showed superior external generalization to holograms of real tissues with arbitrarily defocus distances.

3.2 Network-with-physics strategy

Different from the network-only strategy, in the network-with-physics strategy, either the physical model and neural network are connected in series for phase recovery (physics-connect-network, PcN), or the neural network is integrated into a physics-based algorithm for phase recovery (network-in-physics, NiP), or the physical model or physics-based algorithm is integrated into a neural network for phase recovery (physics-in-network, PiN). A summary of the network-with-physics strategy is presented in Table 3 and is described below.

**Table 3 Summary of network-with-physics strategy**

| Task | Reference | Input | Output | Network | Training dataset | Loss function |
|---|---|---|---|---|---|---|
| physics-connect-network (PcN) | Rivenson et al.[141] | Initial complex field | Pure complex field | CNN and ResNet | Expt.: 100 pairs | $l_2$-norm |
| | Wu et al.[142] | Initial complex field | Pure complex field (in-focus) | U-Net and ResNet | Expt.: 704 pairs | $l_1$-norm |
| | Huang et al.[114] | Initial complex field | Pure complex field | U-Net and Recurrent CNN | Expt.: 208 groups | GAN loss and $l_1$-norm and SSIM |

| | Reference | Input | Output | Network | Training data | Loss function |
|---|---|---|---|---|---|---|
| | Goy et al.[98] | Initial phase | Pure phase | U-Net and ResNet | Expt.: 9,500 pairs | NPCC |
| | Deng et al.[143] | Initial phase | Pure phase | U-Net and ResNet | Expt.: 9,500 pairs | $l_2$-norm and features of VGG |
| | Deng et al.[144] | (i) Initial phase, (ii) LR and HR phase | (i) LR or HR phase, (ii) Phase | U-Net and ResNet (three) | Expt.: 9,500 pairs | NPCC |
| | Kang et al.[145] | Initial phase | Pure phase | U-Net and ResNet | Expt.: 5,000 pairs | NPCC or SSIM |
| | Zhang et al.[146] | Synthetic initial phase and amplitude | HR phase and amplitude | CNN and ResNet | Sim.: 23,040 groups | $l_1$-norm |
| | Moon et al.[147] | initial superimposed phase | Pure phase | U-Net | Expt.: 1,500 pairs | GAN loss |
| Network-in-physics (NiP) | Metzler et al.[148] | Noisy phase | Denoised phase | DnCNN | Sim.: 300,000 pairs | --- |
| | Wu et al.[149] | Noisy phase | Denoised phase | DnCNN | Sim.: --- | --- |
| | Bai et al.[150] | Noisy phase | Denoised phase | DnCNN | Sim.: 300,000 pairs | --- |
| | Wang et al.[151] | Noisy phase | Denoised phase | DnCNN | Sim.: --- | $l_2$-norm |
| | Chang et al.[152] | Noisy phase and amplitude | Denoised phase and amplitude | FFDNet | Sim.: 10,000 pairs | --- |
| | Işıl et al.[153] | Noisy phase | Denoised phase | U-Net | Sim.: 3,000 pairs | $l_2$-norm |
| | Kumar et al.[154] | Noisy phase | Denoised phase | U-Net and ResNet | --- | --- |
| | Jagatap et al.[155,156] | Fixed random vector | Phase | Decoder | Sim.: 1 | $l_2$-norm |
| | Zhou et al.[157] | Fixed random matrix | Phase | SegNet | Sim. and Expt.: 1 | $l_2$-norm |
| | Shamshad et al.[158] | Fixed random matrix | Phase | U-Net | Sim.: 1 | $l_2$-norm |
| | Bostan et al.[159] | Fixed random vector and Zernike polynomials | Phase and aberrations | Decoder or fully connected network | Expt.: 1 | $l_2$-norm |
| | Lawrence et al.[160] | Fixed random vector | Phase | Decoder | Sim.: 1 | Poisson likelihood |
| | Niknam et al.[161] | Fixed random vector | Phase and amplitude | Decoder | Expt.: 1 | $l_2$-norm |
| | Ma et al.[162] | Fixed random vector | Phase | Decoder | Sim.: 1 | $l_2$-norm |
| | Chen et al.[163] | Fixed random vector | Phase, amplitude, pupil aberration and illumination fluctuation factor | Decoders or fully connected networks | Sim.: 1 | $l_2$-norm |
| | Hand et al.[164] | Phase or random vector | Phase | VAE or DCGAN | Sim.: 60,000 and 200,000 pairs | $l_2$-norm |
| | Shamshad et al.[165–167] | Random vector | Phase | DCGAN | Sim.: 60,000 73,257 pairs | $l_2$-norm |
| | Hyder et al.[168] | Random vector | Phase | DCGAN | Sim.: 202,599 pairs | $l_2$-norm |
| | Uelwer et al.[169] | Random vector | Phase | VAE, DCGAN or Style-GAN | Sim.: --- | $l_2$-norm, LPIPS and Wasserstein adversarial loss |
| Physics-in-network (PiN) | Wang et al.[170] | Intensity | Phase | deGEC-SR-Net | Sim.: 100 pairs | $l_2$-norm |
| | Naimipour et al.[171,172] | Intensity | Phase | Auto-encoder network | Sim.: 2,048 pairs | $l_2$-norm |
| | Zhang et al.[173] | Intensity | Phase and amplitude | Complex U-Net (untrained) | Sim.: 1 | $l_2$-norm |
| | Shi et al.[174] | Intensity | Phase | Deep shrinkage network (DSN) | Sim.: 204,800 pairs | $l_2$-norm |
| | Wu et al.[175] | Intensity | Phase | Cascaded CNN | Sim. and Expt.: 400 and 140 pairs | $l_2$-norm |
| | Yang et al.[176] | Intensity | Phase | CNN in space and frequency domain | Sim.: 400 and 60,000 pairs | $l_2$-norm and edge loss |

**Physics-connect-network (PcN).** In this scheme, the role of the neural network is to extract and separate the pure phase from the initial estimate that may suffer from spatial artifacts or low resolution, which allows the neural network to perform a simpler task than the network-only strategy; typically, the initial phase is calculated using a physical model (Fig. 19).

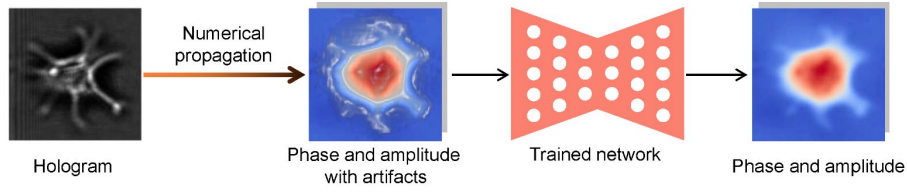

Fig. 19 Description of physics-connect-network (PcN).

Rivenson et al.[141] first applied this PcN scheme in holographic reconstruction in 2018. They used numerical propagation to calculate the initial complex field (including real and imaginary parts) from a single intensity-only hologram, which contained twin-image and self-interference-related spatial artifacts, then used a data-driven trained neural network to extract the pure complex field from the initial estimate. Compared with the axial multi-intensity alternating projection algorithm[33–35], their PcN scheme reduces the number of required holograms by 2-3 times while improving the computation time by more than three times. Wu et al.[142] then extended the depth of field (DOF) based on this work, by training a neural network with pairs of randomly de-focused complex fields and the corresponding in-focus complex field. Meanwhile, Huang et al.[114] proposed to use a recurrent CNN[177] for the PcN scheme and the network-only strategy. They compared the performance of the neural networks with the hologram or initial complex field as input in the same background, and found that the network-only strategy is more robust to sparse samples while the PcN scheme performs better inference on dense samples. Goy et al.[98] applied the PcN scheme to phase recovery under weak-light illumination, which is more ill-posed than conventional phase recovery. They showed that the inference performance of the PcN scheme is stronger than that of the network-only strategy under weak-light illumination, especially for dense samples in the extreme photon level case (1 photon). Further, Deng et al.[143] introduced a default feature perceptual loss of the VGG layer into the loss function for neural network training, which inferred more fine details than that of the NPCC loss function. They also improved the spatial resolution and noise robustness by learning the low-frequency and high-frequency bands, respectively, through two neural networks and synthesizing these two bands into full-band reconstructions with a third neural network[144]. By introducing random phase modulation, Kang et al.[145] further improved the phase recovery ability of the PcN scheme under weak-light illumination. Zhang et al.[146] extended the PcN scheme to FP, inferring high-resolution phase and amplitude using the initial phase and amplitude synthesized from the intensity images as input to a neural network. Moon et al.[147] extended the PcN scheme to off-axis holography, using numerical propagation to obtain the initial phase from the Gaber hologram as the input to the neural network.

**Network-in-physics (NiP).** Regarding phase recovery as one of the most general optimization problems, this approach can be expressed as

$$\arg\min_{\theta}\|I_x - H(\theta)\|_2^2 + R(\theta), \tag{4}$$

where $H(\cdot)$ is the physical model (such as numerical propagation, Fourier transform, or FP measurement model), $\theta$ is the phase, $I_x$ is the measured intensity image of a test sample, and $R(\theta)$ is a regularized constraint. According to the Regularization-by-Denoising (RED)[178] framework, a pre-trained neural network for denoising can be used as the regularized constraint:

$$\arg\min_{\theta}\|I_x - H(\theta)\|_2^2 + R(\theta), \quad \text{with } R(\theta) = \lambda\theta^T(\theta - D(\theta)), \tag{5}$$

where $D(\theta)$ is a pre-trained neural network for denoising, and $\lambda$ is a weight factor to control the strength of regularization. Metzler et al.[148] used the above algorithm for phase recovery and called it PrDeep. They used a DnCNN trained on 300,000 pairs of data as a denoiser and FASTA[179] as a solver. In comparison with other conventional iterative methods, PrDeep demonstrates excellent robustness to noise. Wu et al.[149] proposed an online extension of PrDeep, which adopts the online processing of data by using only a random subset of measurements at a time. Bai et al.[150] extended PrDeep to incorporate a contrast-transfer-function-based forward operator in $H(\cdot)$ for phase recovery. Wang et al.[151] improved PrDeep by changing the solver from FASTA to ADMM, which further improved the noise robustness. Chang et al.[152] used a generalized-alternating-projection solver to further expand the performance of PrDeep and made it suitable for the recovery of complex fields. Işıl et al.[153] embedded a trained neural network denoiser into HIO, removing artifacts from the results after each iteration. On this basis, Kumar et al.[154] added total-variation prior together with the denoiser for regularization.

In addition, according to the deep image prior (DIP)[180,181], even an untrained neural network itself can be used as a structural prior for regularization (Fig. 20):

$$\arg\min_{\omega}\left\|I_x - H\left(g_\omega(z_f)\right)\right\|_2^2, \tag{6}$$

where $g_\omega(\cdot)$ is an untrained neural network with trainable parameters $\omega$ that usually takes a generative decoder architecture, $I_x$ is the measured intensity image of a test sample, and $z_f$ is a fixed random vector as latent code.

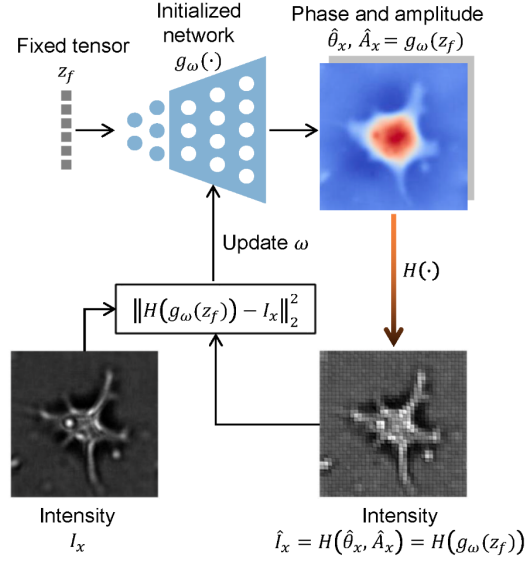

Fig. 20 Description of DIP-based phase recovery.

This DIP-based approach was first introduced to phase recovery by Jagatap et al.[155]. They solved Eq. (6) using the gradient descent and projected gradient descent algorithms by optimizing over trainable parameters $\omega$, both of which outperform sparse truncated amplitude flow (SPARTA) algorithm. In follow-up work, they provided rigorous theoretical guarantees for the convergence of their algorithm[156]. Zhou et al.[157] applied this DIP-based algorithm to ODT, alleviating the effects of the missing cone problem. Shamshad et al.[158] extended this DIP-based algorithm to subsampled FP, achieving better reconstructions at low subsampling ratios and high noise perturbations. In order to make the algorithm adaptive to different aberrations, Bostan et al.[159] added a fully connected neural network with Zernike polynomials as the fixed input, and used it as the second structural prior. In the holographic setting with a reference beam, Lawrence et al.[160] demonstrated the powerful information reconstruction ability of the DIP-based algorithm in extreme cases such as low photon counts, beamstop-obscured frequencies, and small oversampling. Niknam et al.[161] used the DIP-based algorithm to recover complex fields from an in-line hologram. They further improved the twin-image artifacts suppression capability through some additional regularization, such as bounded activation function, weight decay, and parameter perturbation. Ma et al.[162] embed an untrained generation network into the ADMM algorithm to solve the phase recovery at low subsampling ratios, and achieved better results than the gradient descent and projected gradient descent algorithms of Jagatap et al.[155]. Chen et al.[163] extended the DIP-based algorithm to FP, in which four parallel untrained neural networks were used for generating phase, amplitude, pupil aberration, and illumination fluctuation factor correction, respectively.

Similarly, a pre-trained generative neural network can also be used as a generative prior, assuming that the target phase is in the range of the output of this trained neural network (Fig. 21):

$$\arg \min_{z} \|I_x - H(G(z))\|_2^2, \quad (7)$$

where $G(\cdot)$ is a pre-trained fixed neural network that usually takes a generative decoder architecture, $I_x$ is the measured intensity image of a test sample, and $z$ is a latent code tensor to be searched. Due to the use of the generative neural network, the multi-dimensional phase that originally needed to be iteratively searched is converted into a low-dimensional tensor, and the solution space is also limited within the range of the trained generative neural network.

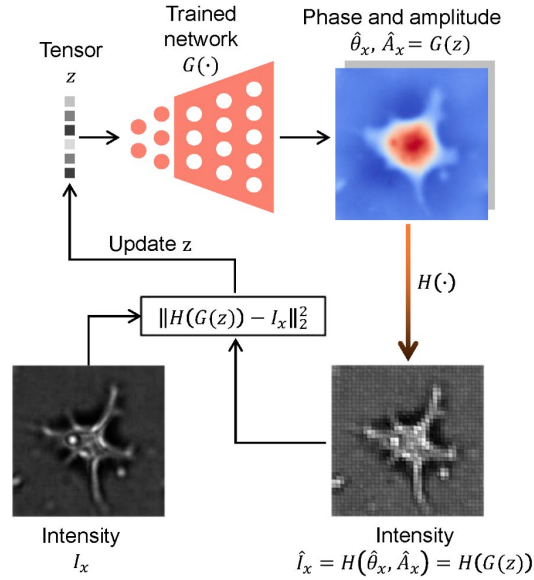

Fig. 21 Description of generative-prior-based phase recovery.

Hand et al.[164] used generative prior for phase recovery with rigorous theoretical guarantees for random Gaussian measurement matrix, showing better performance than SPARTA at low subsampling ratios. Later on, Shamshad et al.[165] experimentally verified the robustness of the generative-prior-based algorithm to low subsampling ratios and strong noise in the coded diffraction setup. Then, Shamshad et al.[166] extended this generative-prior-based algorithm to subsampled FP. Hyder et al.[168] improved over this by combining the gradient descent and projected gradient descent methods with AltMin-based non-convex optimization methods. As a general defect, the trained generative neural network will limit the solution space to a specific range related to the training dataset, so that the iterative algorithm cannot search beyond this range. Therefore, Shamshad et al.[167] set both the input and previously fixed parameters of the trained generative neural network to be trainable. As another solution, Uelwer et al.[169] extended the range of the trained generative neural network by intermediate layer optimization.

**Physics-in-network (PiN).** According to the algorithm unrolling/unfolding technique proposed by Gregor and LeCun[182], physics-based iterative algorithms can be unrolled as an interpretable neural network architecture (Fig. 22). Wang et al.[170] unrolled an algorithm called decentralized generalized expectation consistent signal recovery (deGEC-SR) into a neural network with trainable parameters, which exhibits stronger robustness using fewer iterations than the original deGEC-SR. Naimipour et al.[171,172] used the algorithm unrolling technique in reshaped Wirtinger flow and SPARTA. Zhang et al.[173] unrolled the iterative process of the alternative projection algorithm into complex U-Nets. Shi et al.[174] used a deep shrinkage network and dual frames to unroll the proximal gradient algorithm in coded diffraction imaging. Wu et al.[175] integrated the Fresnel forward operator and TIE inverse model into a neural network, which can be efficiently trained with a small number of datasets and is suitable for transfer learning. Yang et al.[176] unrolled the classic HIO algorithm into a neural network that combines information both in the spatial domain and frequency domain.

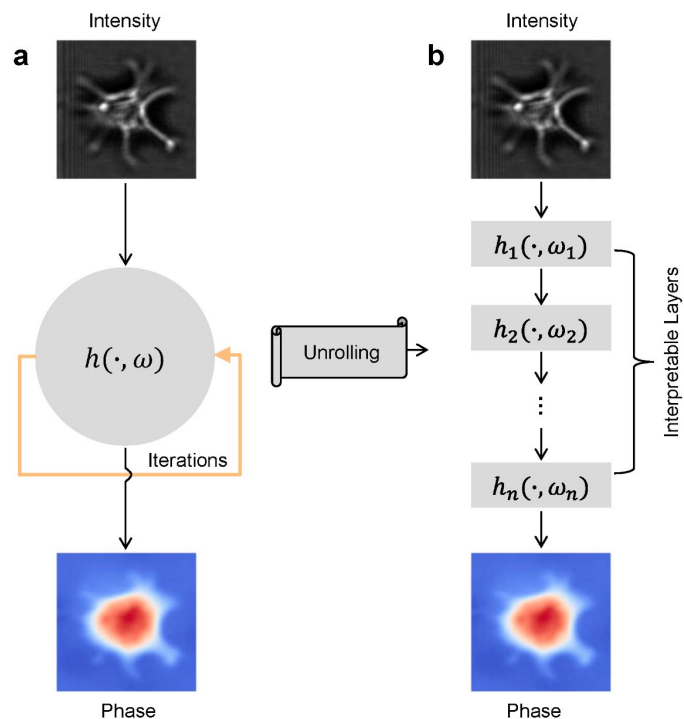

Fig. 22 Description of physics-in-network (PiN). **a** A physics-based iterative algorithm. **b** A corresponding unrolled neural network. The iteration step $h$ with algorithm parameters $\omega$ in (a) is unrolled and transferred to the network layers $h_1$, $h_2$,..., $h_n$ with network parameters $\omega_1$, $\omega_2$,..., $\omega_n$ in (b). The unrolled neural network is trained with the dataset in an end-to-end manner.

## 4. DL-post-processing for phase recovery

A summary of "DL-post-processing for phase recovery" is presented in Table 4 and is described below, including noise reduction (Section 4.1), resolution enhancement (Section 4.2), aberration correction (Section 4.3), and phase unwrapping (Section 4.4).

**Table 4 Summary of "DL-post-preprocessing for phase recovery"**

| Task | Reference | Input | Output | Network | Training dataset | Loss function |
|---|---|---|---|---|---|---|
| Noise reduction | Jeon et al.[183] | Noisy hologram | Noise-free hologram | U-Net | Sim.: 384,000 pairs | $l_2$-norm and Edge |
| | Choi et al.[184] | Noisy tomogram | Noise-free tomogram | U-Net | Expt.: 455 and 5,057 (unpaired) | Cycle-GAN loss |
| | Zhang et al.[185] | Noisy wrapped phase | Noise-free wrapped phase | CNN | Sim.: 500 pairs | --- |
| | Yan et al.[186,187] | Noisy sine and cosine | Noise-free sine and cosine | ResNet | Sim.: 40,000 and 30,000 pairs | $l_2$-norm |
| | Montresor et al.[188] | Noisy sine and cosine | Noise-free sine and cosine | DnCNN | Sim.: 40 pairs (15,360 patches) | $l_2$-norm |
| | Tahon et al.[189,190] | Noisy sine and cosine | Noise-free sine and cosine | DnCNN | Sim.: 25 pairs and 128 pairs | $l_2$-norm |
| | Fang et al.[191] | Noisy real, imaginary | Noise-free real, imaginary | U-Net | Sim.: 4,000 pairs | GAN loss |
| | Murdaca et al.[192] | Noisy real, imaginary, and amplitude | Noise-free real, imaginary, and amplitude | U-Net | Sim.: 5,400 pairs | $l_2$-norm |
| | Tang et al.[193] | Fixed noise matrix | Noise-free phase | U-Net (untrained) | Expt.: 1 | $l_2$-norm, gradient, and variance |
| Resolution enhancement | Liu et al.[194,195] | LR phase and amplitude | HR phase and amplitude | U-Net | Expt.: >50,000 pairs | GAN loss |
| | Jiao et al.[196] | LR phase from DPM | HR phase from SLIM | U-Net | Expt.: >1,200 pairs (>100 cells) | $l_2$-norm |
| | Butola et al.[197] | LR phase | HR phase | U-Net | Expt.: 2,355 pairs and 2,279 pairs | GAN loss |
| | Meng et al.[198] | LR phase from SI-DHM | HR phase from SI-DHM | U-Net | Expt.: 3,800 pairs | $l_2$-norm |
| | Li et al.[199] | LR phase from qDPC | HR phase from qDPC | U-Net | Expt.: 1,680 pairs | $l_2$-norm |
| | Gupta et al.[200] | LR phase | HR phase | U-Net | Expt.: 700-2,000 (unpaired) | Cycle-GAN loss |
| | Lim et al.[201] | LR 3D RI tomogram | HR 3D RI tomogram | Residual 3D U-Net | Sim.: 1,600 pairs | $l_2$-norm |
| | Ryu et al.[202] | LR 3D RI tomogram | HR 3D RI tomogram | 3D U-Net | Expt.: 217 and 614 pairs | $l_2$-norm |
| Aberration correction | Nguyen et al.[203] | Phase | Binary segmentation | U-Net | Expt.: 1,836 pairs | Cross entropy |
| | Ma et al.[204] | Hologram | Binary segmentation | U-Net | Expt.: 1,000 pairs | Cross entropy |
| | Lin et al.[205] | Phase and its gradient | Binary segmentation | U-Net and ResNet | Expt.: 1,800 pairs | Dice loss |
| | Xiao et al.[206] | Phase | Zernike coefficient | CNN | Expt.: 10,000 pairs | $l_2$-norm |
| | Zhang et al.[207] | Aberrated intensity and phase | Phase | U-Net | Sim.: >10,000 groups | $l_2$-norm or $l_1$-norm |
| | Tang et al.[208] | Fixed tensor | Zernike coefficient | MLP (untrained) | Expt. and Sim.: 1 | $l_2$-norm and sparse constraints |
| unwrapping | Dardikman et al.[209,210] | Wrapped phase | Unwrapped phase | ResNet | Sim.: 7,936 pairs | $l_2$-norm |

| | | | | | |
|---|---|---|---|---|---|
| Wang et al.[211] | Wrapped phase | Unwrapped phase | U-Net and ResNet | Sim.: 30,000 pairs | $l_2$-norm |
| He et al.[212] | Wrapped phase | Unwrapped phase | 3D-ResNet | Expt.: --- | --- |
| Ryu et al.[213] | Wrapped phase | Unwrapped phase | ReNet | Expt. and Sim.: --- | Total variation and variance |
| Dardikman et al.[214] | Wrapped phase | Unwrapped phase | ResNet | Expt.: 7,500 pairs | $l_2$-norm |
| Qin et al.[215] | Wrapped phase | Unwrapped phase | U-Net and ResNet | Sim.: 30,000 pairs | $L_1$-norm |
| Perera et al.[216] | Wrapped phase | Unwrapped phase | U-Net and LSTM | Sim.: 6,000 pairs | Total variation and variance |
| Park et al.[217] | Wrapped phase | Unwrapped phase | U-Net | Expt.: 5,200 pairs | GAN loss |
| Zhou et al.[218] | Wrapped phase and wrap count | Unwrapped phase | U-Net and EfficientNet | Sim.: 6,000 pairs | $l_1$-norm and residual |
| Xu et al.[219] | Wrapped phase | Unwrapped phase | U-Net | Sim.: 6,000 pairs | SSIM |
| Zhou et al.[220] | Wrapped phase | Unwrapped phase | U-Net | Sim.: 158 and 1,036 pairs | GAN loss |
| Xie et al.[221] | Wrapped phase | Unwrapped phase | U-Net | Sim.: 17,000 pairs | $l_2$-norm |
| Zhao et al.[222] | Wrapped phase and weighted map | Unwrapped phase | U-Net and ResNet | Sim.: 22,500 pairs | $l_1$-norm |
| Liang et al.[223] | Wrapped phase | Wrap count | --- | --- | --- |
| Spoorthi et al.[224] | Wrapped phase | Wrap count | SegNet | Sim.: 10,000 pairs | Cross entropy |
| Spoorthi et al.[225] | Wrapped phase | Wrap count | SegNet and DenseNet | Sim.: 30,000 pairs | Cross entropy and residue and $l_1$-norm |
| Zhang and Liang et al.[185,226] | Wrapped phase | Wrap count | U-Net | Sim.: 9,500 pairs | Cross entropy |
| Zhang et al.[227] | Wrapped phase | Wrap count | DeepLab-V3+ | Sim.: 25,000 pairs | Cross entropy |
| Zhu et al.[228] | Wrapped phase | Wrap count | DeepLab-V3+ | Sim.: 20,000 pairs | Cross entropy |
| Wu et al.[229] | Wrapped phase | Wrap count | U-Net and FRRNet | Sim.: 12,000 pairs | Cross entropy |
| Zhao et al.[230] | Wrapped phase | Wrap count | ResNet | Sim.: 22,000 pairs | Cross entropy |
| Vengala et al.[231,232] | Wrapped phase | Wrap count and denoised wrapped phase | Y-Net | Sim.: 2,000 pairs | Cross entropy and $l_2$-norm |
| Zhang et al.[233] | Wrapped phase | Wrap count | U-Net and ASPP and PSA | Sim.: 10,000 pairs | Weighted cross entropy |
| Huang et al.[234] | Wrapped phase | Wrap count | HRNet | Sim.: 30,000 pairs | Cross entropy |
| Wang et al.[235] | Wrapped phase | Wrap count | U-Net, ASPP and EEB | Sim.: --- | Cross entropy |
| Zhou et al.[236] | Wrapped count | Wrap count gradient | CNN | Sim.: 52,391 pairs | Cross entropy |
| Wang et al.[237] | Wrapped count and quality map | Wrap count gradient | U-Net | Sim.: 164,726 pairs | Cross entropy and dice loss |
| Sica et al.[238] | Wrapped count | Wrap count gradient | U-Net | Sim.: >70,000 pairs | Cross entropy, Jaccard distance, and $l_1$-norm |
| Li et al.[239] | Wrapped count | Wrap count gradient | U-Net and ResNet | Sim.: 14,100 pairs | Cross entropy |
| Wu et al.[240,241] | Wrapped count | Discontinuity map | CNN and ASPP | Sim.: 8,000 pairs | $l_2$-norm, cross entropy, and dice loss |
| Zhou et al.[242] | Residue image | Branch-cut map | CNN | Sim.: 26,928 pairs | Cross entropy |

## 4.1 Noise reduction

In addition to being part of the pre-processing in Section 2.2, noise reduction can also be performed after phase recovery (Fig. 23). Jeon et al.[183] applied the U-Net to perform speckle noise reduction on digital holographic images in an end-to-end manner. Their deep learning method takes only 0.92 s for a reconstructed hologram of 2048×2048, while other conventional methods take tens of seconds because of the requirement of multiple holograms. Choi et al.[184] introduced the cycle-GAN to train neural networks for noise reduction by unpaired datasets. They demonstrated the advantages of this un-paired-data-driven method with tomograms of different cell samples in optical diffraction chromatography: the non-data-driven ways either remove coherent noise by blurring the entire images or perform no effective denoising, whereas their method can simultaneously remove the noise and preserve the features of the sample.

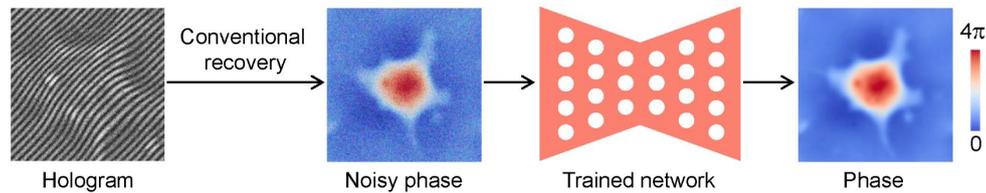

Fig. 23 Description of deep-learning-based phase noise reduction.

Zhang et al.[185] first proposed to suppress noise directly on the wrapped phase via a neural network. However, this direct way may lead to many wrong jumps in the wrapped phase, which results in larger errors in the unwrapped phase. Thus, Yan et al.[186,187] proposed to do noise reduction on the sine and cosine (numerator and denominator) images of the phase via a neural network, and then calculated the wrapped phase from denoised sine and cosine images by the arctangent function. Almost simultaneously, Montresor et al.[188] introduced the DnCNN into speckle noise reduction for phase data by their sine and cosine images. As it is difficult to simultaneously collect the phase data with and without speckle noise in an experimental manner, they used a simulator based on a double-diffraction system to numerically generate the dataset. Furthermore, their method yields comparable standard deviation to the WFT and better peak-to-valley, while costing less time. Building on this work, Tahon et al.[189] designed a dataset (HOLODEEP) for speckle noise reduction in soft conditions and used a shallower network for faster inference. To go further, they released a more comprehensive dataset for conditions of severe speckle noise[190]. Fang et al.[191] applied GAN to do speckle noise reduction for phase. Murdaca et al.[192] applied this deep-learning-based phase noise reduction to interferometric synthetic aperture radar (InSAR)[243]. The difference is that in addition to the sine and cosine images of the phase, the neural network also reduces noise for the amplitude images at the same time. Tang et al.[193] proposed to iteratively reduce the coherent noise in phase with an untrained U-Net.

4.2 Resolution enhancement

Similar to Section 2.1, resolution enhancement can also be performed after phase recovery as post-processing (Fig. 24). Liu et al.[194,195] first used a neural network to infer the corresponding high-resolution phase from the low-resolution phase. They trained two GANs with both a pixel super-resolution system and a diffraction-limited super-resolution system, which was demonstrated on biological thin tissue slices with the analysis of spatial frequency spectrum. Moreover, they pointed out that this idea can be extended to other resolution-limited imaging systems, such as using a neural network to build a passageway from off-axis holography to in-line holography. Later, Jiao et al.[196] proposed to infer the high-resolution noise-free phase from an off-axis-system-acquired low-resolution version with a trained U-Net. To collect the paired dataset, they developed a combined system with diffraction phase microscopy (DPM)[244] and spatial light interference microscopy (SLIM)[22] to generate both holograms from the same field of view. After training, the U-Net retains the advantages of both the high acquisition speed of DPM and the high transverse resolution of SLIM.

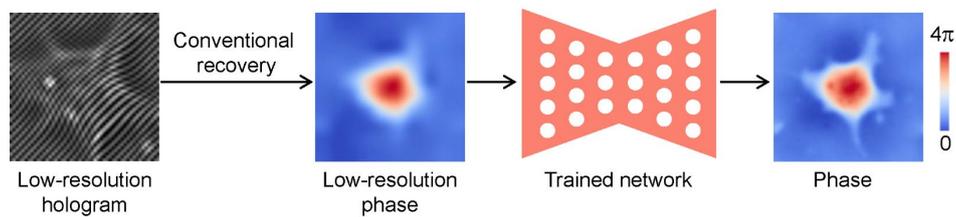

Fig. 24 Description of deep-learning-based phase resolution enhancement.

Subsequently, Butola et al.[197] extended this idea to partially spatially coherent off-axis holography, where the phase recovered at low-numerical-apertures objectives was used as input, and the phase recovered at high-numerical-apertures objectives was used as ground-truth. Since low-numerical-apertures objectives have a larger field of view, they aim to obtain a higher resolution at a larger field of view, i.e., a higher spatial bandwidth product. Meng et al.[198] used structured-illumination digital holographic microscopy (SI-DHM)[245] to collect the high-resolution phase as ground-truth. To supplement more high-frequency information by two cascaded neural networks, they used the low-resolution phase along with the high-resolution amplitude inferred from the first neural network both as inputs of the second neural network. Subsequently, Li et al.[199] extended this resolution-enhanced post-processing method to quantitative differential phase-contrast (qDPC)[246] imaging for high-resolution phase recovery from the least number of experimental measurements. To solve the problem of out-of-memory for the large size of the input, they disassembled the full-size input into some sub-patches. Moreover, they found that the U-Net trained on the paired dataset has a smaller error than the paired GAN and the unpaired GAN. While for GAN, there is more unreasonable information in the inferred phase, which is absent in the ground-truth. Gupta et al.[200] took advantage of the high spatial bandwidth product of this method to achieve a classification throughput rate of 78,000 cells per second with an accuracy of 76.2%.

For ODT, due to the limited projection angle imposed by the numerical aperture of the objective lens, there are certain spatial frequency components that cannot be measured, which is called the missing cone problem. To address this problem via a neural network, Lim *et al.*[201] and Ryu *et al.*[202] built a 3D RI tomogram dataset for 3D U-Net training, in which the raw RI tomograms with poor axial resolution were used as input, and the resolution-enhanced RI tomograms from the iterative total variation algorithm were used as ground-truth. The trained 3D U-Net can infer the high-resolution version directly from the raw RI tomograms. They demonstrated the feasibility and generalizability using bacterial cells and a human leukemic cell line. Their deep-learning-based resolution-enhanced method outperforms conventional iterative methods by more than an order of magnitude in regularization performance.

4.3 Aberration correction

For holography, especially in the off-axis case, the lens and the unstable environment of the sample introduce phase aberrations superimposing on the phase of the sample. To recover the pure phase of the sample, the unwanted phase aberrations should be eliminated physically or numerically. Physical approaches compensate for the phase aberrations by recovering the background phase without the sample from anther hologram, which requires more setups and adjustments[247,248].

As for numerical approaches, the compensation of the phase aberrations can be directly achieved by Zernike polynomial fitting (ZPF)[249] or principal-component analysis (PCA)[250]. Yet, in these numerical methods, the aberration is predicted from the whole phase, where the object area should not be considered as an aberration. Thus, before using the Zernike polynomial fitting, the neural network can be used to find out the object area and the background area to avoid the influence of the background area and improve the compensation effect (Fig. 25). This segmentation-based idea, namely CNN+ZPF, was first proposed by Nguyen *et al.*[203] in 2017. They manually made binary masks as ground-truth for each phase to distinguish the area of the background and sample. After comparison on different real samples, they found that the compensated result of the CNN+ZPF contains flatter background than that of PCA. However, the aberration in the initial phase makes it more difficult to do segmentation from the already weak phase distribution of the boundary features, especially for the large tilted phase aberrations. To address this problem, Ma *et al.*[204] proposed to do segmentation with hologram instead of phase as neural network input. Lin *et al.*[205] applied the CNN+ZPF to real-time phase compensation with a phase-only SLM.

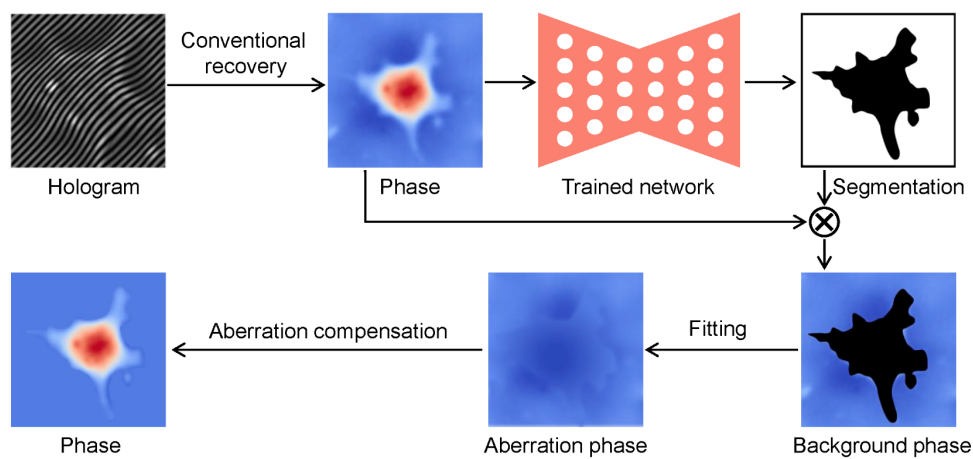

Fig. 25 Description of deep-learning-based phase aberration correction.

In addition to the way of CNN+ZPS, Xiao *et al.*[206] directly inferred the Zernike coefficient of aberration from the initial phase via a neural network, which costs less computation. They trained a neural network specifically for bone cells, and used this efficient method to achieve long-term morphological observation of living cells. Zhang *et al.*[207] used a trained neural network to infer the in-focus phase from the de-focus aberrated intensity and phase. Tang *et al.*[208] introduced the sparse constraint into the loss function and iteratively inferred the corresponding phase aberrations from the initial phase or fixed tensor with an untrained neural network and Zernike model.

4.4 Phase unwrapping

In the interferometric and optimization-based phase recovery methods, the recovered light field is in the form of complex exponential. Hence the calculated phase is limited in the range of $(-\pi, \pi]$ on account of the arctangent function. Therefore, the information of the sample cannot be obtained unless the absolute phase is first estimated from the wrapped phase, the so-called phase unwrapping. In addition to phase recovery, the phase unwrapping problem also arises in magnetic resonance imaging[251], fringe projection profilometry[252], and InSAR. Most conventional methods are based on the phase continuity assumption, and some cases, such as noise, breakpoints, and aliasing, all violate the Itoh condition and affect the effect of the conventional methods[253]. The advent of deep learning has made it possible to perform phase unwrapping in the above cases. According to the different use of the neural network, these deep-learning-based phase unwrapping methods can be divided into the following three categories (Fig. 26)[48]. Deep-learning-performed regression method (dRG) estimates the absolute phase directly from the wrapped phase by a neural network (Fig. 26a)[209–222]. Deep-learning-performed wrap count method (dWC) first estimates the wrap count from the wrapped phase by a neural network, and then calculates the absolute phase from the wrapped phase and the estimate wrap count (Fig. 26b)[185,223–233]. Deep-learning-assisted method (dAS) first estimates the wrap count gradient or discontinuity from the wrapped phase by a neural network; next, either reconstruct the wrap count from the wrap count gradient and then calculate the absolute phase like dWC[238,239], or directly use optimization-based or branch-cut algorithms to obtain the absolute phase from the warp count gradient or the discontinuity (Fig. 26c)[236,237,240–242].

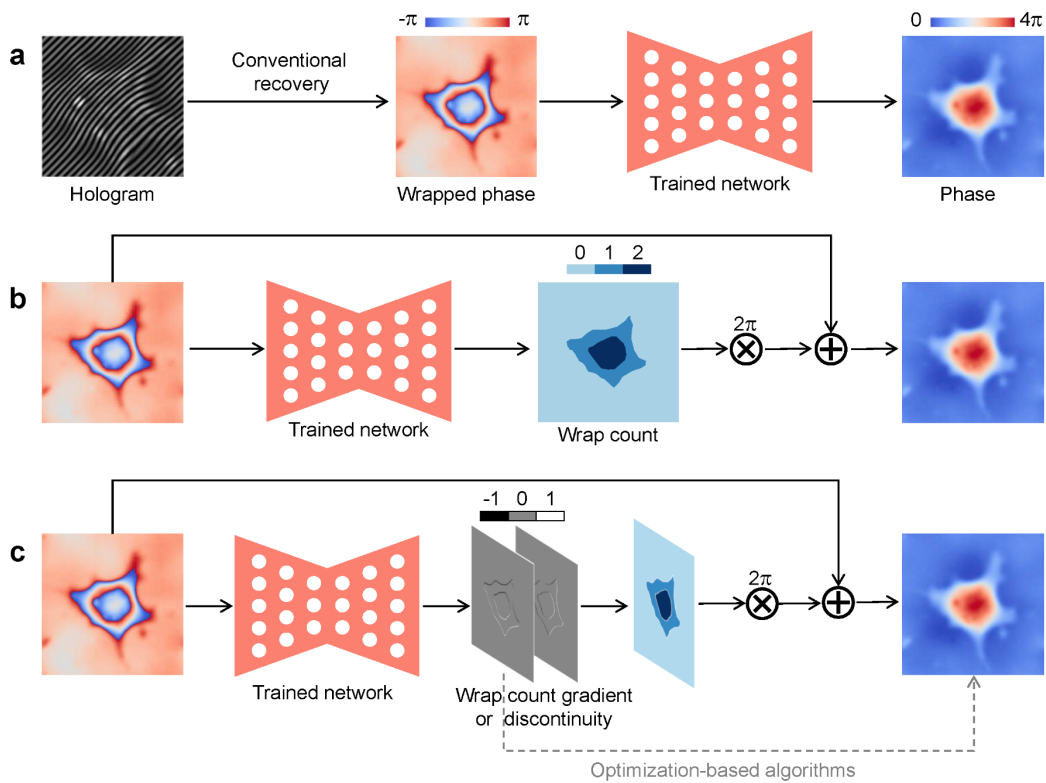

**Fig. 26 Description of deep-learning-based phase unwrapping. a** Deep-learning-performed regression method. **b** Deep-learning-performed wrap count method. **c** Deep-learning-assisted method.

**Deep-learning-performed regression method (dRG).** Dardikman *et al.*[209] presented the dRG method, which utilizes a residual-block-based CNN with a dataset of simulated steep cells. They also validated the dRG method post-processed by congruence in actual cells and compared it with the performance of the dWC method[210]. Then, Wang *et al.*[211] introduced the U-Net and a phase simulation generation method into the dRG method, wherein they evaluated the trained network on real samples, examined the network's generalization ability through middle-layer visualization, and demonstrated the superiority of the dRG method over conventional methods in noisy and aliasing cases. In the same year, He et al.[212] and Ryu et al.[213] evaluated the ability of the 3D-ResNet and recurrent neural network (ReNet) to perform phase unwrapping using magnetic resonance imaging data. Dardikman et al.[214] released their real sample dataset as open-source. They demonstrated that the congruence could enhance the accuracy and robustness of the dRG method, particularly when dealing with a limited number of wrap count. Qin et al.[215] utilized a Res-UNet with a larger capacity to achieve higher accuracy and introduced two new evaluation indices. Perera et al.[216] and Park et al.[217] introduced the long short-term memory (LSTM) network and GAN into phase unwrapping. Zhou et al.[218,254] enhanced the robustness and efficiency of the dRG method by doing preprocessing and postprocessing steps for the U-Net with EfficientNet[254] backbone. Xu et al.[219] improved the accuracy and robustness of the U-Net by adding more middle-layers and skip connections and using a composite loss function. Zhou et al.[220] used the GAN in the InSAR phase unwrapping and avoided the blur in the unwrapped phase by combining the $l_1$ loss and adversarial loss. Xie et al.[221] trained four networks for different noise levels, which made each network more focus on a specific noise level. Zhao et al.[222] added a weighted map as the prior to the neural network to make it more focused on the area near the jump edge, similar to an additional attention mechanism. Different from the above methods, Vithin et al.[255,256] proposed to use the Y-Net[92] to infer the phase gradients from a wrapped phase and then calculate the absolute phase.

**Deep-learning-performed wrap count method (dWC).** Liang et al.[223] and Spoorthi et al.[224] first proposed this idea in 2018. Spoorthi et al.[224] proposed a phase dataset generation method by adding and subtracting Gaussian functions with randomly varying mean and variance values, and used the clustering-based smoothness to alleviate the classification imbalance of the SegNet. Further, the prediction accuracy of their methods was improved by introducing the prior of absolute phase values and gradients into the loss function, which they called Phase-Net2.0[225]. Zhang and Liang et al.[226,185] sequentially used three networks to perform phase unwrapping by wrapped phase denoising, wrap count predicting, and post-processing. In addition, they proposed to generate a phase dataset by weighted adding Zernike polynomials of different orders. Immediately after, Zhang and Yan et al.[227] verified the performance of the network DeepLab-V3+, but the resulting wrap count still contained a small number of wrong pixels, which will propagate error through the whole phase maps in the conventional phase unwrapping process. They thus proposed to use refinement to correct the wrong pixels. To further improve the unwrapped phase, Zhu et al.[228] proposed to use the median filter for the second post-processing to correct wrong pixels in the wrap count predictions. Wu et al.[229] enhanced the simulated phase dataset by adding the noise from real data. They also used the full-resolution residual network (FRRNet) with U-Net to further optimize the performance of the U-Net in the Doppler optical coherence tomography. By comparison with real data, their proposed network holds a higher accuracy than that of the Phase-Net and DeepLab-V3+. As for applying the dWC to point diffraction interferometer, Zhao et al.[230] proposed an image-analysis-based post-processed method to alleviate the classification imbalance of the task and adopted the iterative-closest-point stitching method to realize dynamic resolution. Vengala et al.[92,231,232] used the Y-Net[92] to reconstruct the wrap count and pure wrapped phase at the same time. Zhang et al.[233] added atrous spatial pyramid pooling (ASPP), positional self-attention (PSA), and edge-enhanced block (EEB) to the U-Net to get higher accuracy and stronger robustness than the networks used in the above methods. Huang et al.[234] applied the HRNet to the dWC methods. Their method still needs the median filter for post-processing, although the performance is better than that of the Phase-Net and DeepLab-V3+. Wang et al.[235] proposed another EEB based on Laplacian and Prewitt edge enhancement operators for the network, which further enhances classification accuracy and avoids the use of post-processing.

**Deep-learning-assisted method (dAS).** The conventional methods estimate the wrap count gradient under the phase continuity assumption, which hence is disturbed by disadvantages factors such as noise. To get rid of it, Zhou et al.[236] proposed to estimate the wrap count gradient via a neural network instead of conventional methods. Since the noisy wrapped phase and the corresponding correct wrap count gradient are used as training datasets, the trained neural network is able to estimate the correct wrap count gradient from the noisy wrapped phase without being limited by the phase continuity assumption. The correct result can be obtained by minimizing the difference between the unwrapped phase gradients and the network-output wrap count gradient. Further, Wang et al.[237] proposed to input a quality map, as the prior, together with the wrapped phase into the neural network to improve the accuracy of the estimated wrap count gradient. Almost simultaneously, Sica et al.[238] directly reconstructed the wrap count from the network-output wrap count gradient and then calculated the absolute phase, like dWC. On this basis, Li et al.[239] improved neural network estimation efficiency by using a single fusion gradient instead of the vertical and horizontal gradients. In addition to estimating the wrap count gradient via a neural network, Wu et al.[240,241] chose to estimate the horizontal and vertical discontinuities with a neural network, and recover the absolute phase by the optimization-based algorithms. Instead of using the wrapped phase as the network input, Zhou et al.[242] embedded the neural network into the branch-cut algorithm to predict the branch-cut map from the residual image, which reduced the computational cost of the branch-cut algorithm.

## 5. Deep learning for phase processing

A summary of "Deep learning for phase processing" is presented in Table 5 and is described below, including segmentation (Section 5.1), classification (Section 5.2), and imaging modal transformation (Section 5.3).

**Table 5 Summary of "Deep learning for phase processing"**

| Task | Reference | Input | Output | Network | Training dataset | Loss function |
|---|---|---|---|---|---|---|
| Segmentation | Yi et al.[257] | Phase of red blood cells | Segmentation map | FCN | Expt.: 35 pairs | --- |
| | Ahmadzadeh et al.[258] | Phase of cardiomyocyte | Segmentation map | FCN | Expt.: 2,000 pairs | --- |
| | Kandel et al.[259] | Phase of sperm cells | Segmentation map | U-Net | Expt.: --- | Cross entropy |
| | Goswami et al.[260] | Phase of virus particles | Segmentation map | U-Net | Expt.: 1,000 pairs | Cross entropy |
| | Hu et al.[261] | Phase of ovary cells | Segmentation map | U-Net and EfficientNet | Expt.: 1,536 pairs | Focal loss and dice loss |
| | He et al.[262] | Phase of HeLa cells | Segmentation map | U-Net and EfficientNet | Expt.: 2,046 pairs | focal loss and dice loss |
| | Zhang et al.[263] | Phase of tissue slices | Segmentation map | mask R-CNN | Expt.: 196 pairs | Cross entropy |
| | Jiang et al.[264] | Phase and amplitude | Segmentation map | DeepLab-V3+ | Expt.: 1,500 pairs | Cross entropy |
| | Lee et al.[265] | 2D RI tomogram | Segmentation map | U-Net | Expt.: 934 pairs | Cross entropy |
| | Choi et al.[266] | 3D RI tomogram | Segmentation map | 3D U-Net | Expt.: 105 pairs | Cross entropy and dice loss |
| Classification | Jo et al.[267] | Phase of cells | Classification | CNN | Expt.: --- | Cross entropy |
| | Karandikar et al.[268] | Phase of cells | Classification | CNN | Expt.: 300 | Cross entropy |
| | Zhang et al.[269] | Phase of tissue slices | Classification | VGG | Expt.: 1,660 | Cross entropy |
| | Butola et al.[270] | Phase of sperm cells | Classification | CNN | Expt.: 10,163 | Cross entropy |
| | Li et al.[271] | Phase of cells | Classification | AlexNet | Expt.: 272 | Cross entropy |
| | Shu et al.[272] | Phase of cells | Classification | Cascaded ResNet | Expt.: 1,521 | Cross entropy |
| | Pitkäaho et al.[273] | Phase and manual feature | Classification | CNN | Expt.: 2,451 | --- |
| | O'Connor et al.[274] | Transfer-learning and manual feature from phase | Classification | LSTM | Expt.: 303 | --- |
| | O'Connor et al.[275] | Transfer-learning and manual feature from phase | Classification for COVID-19 | LSTM | Expt.: 1,474 | --- |
| | Ryu et al.[276] | 3D RI tomogram | Classification (2 and 5 types) | 3D CNN | Expt.: 1,782 | Cross entropy |
| | Kim et al.[277] | 3D RI tomogram | Classification (19 types) | 3D CNN | Expt.: 10,556 | Cross entropy |
| | Wang et al.[278] | Time-lapse amplitude and phase | Classification (3 types) | Pseudo-3D DensNet | Expt.: 16,309 | Cross entropy |
| | Liu et al.[279] | Time-lapse phase | Classification | Pseudo-3D DensNet | Expt.: 5,622 | Cross entropy |
| | Ben Baruch et al.[280] | Phase and spatio-temporal fluctuation map | Classification | ResNet | Expt.: 216 videos | Cross entropy |
| | Singla et al.[281] | Phase of three wavelengths | Classification | CNN | Expt.: 16,200 | --- |
| | Işıl et al.[282] | Phase and amplitude of three wavelengths | Classification | DensNet | Expt.: 33,768 | Cross entropy |
| | Pitkäaho et al.[283] | Phase and amplitude | Classification | CNN | Expt.: --- | --- |

| | Lam et al.[284–286] | Phase and amplitude | Classification | CNN | Sim.: >1,000 Expt.: 4,000 | --- |
|---|---|---|---|---|---|---|
| | Terbe et al.[287] | Phase and amplitude in different defocus distances | Classification (7 types) | 3D ResNet | Expt.: >9,000 | Cross entropy |
| | Wu et al.[288] | Real and imaginary | Classification (5 types) | ResNet | Expt.: 7,000 | Cross entropy |
| Imaging modal transformation | Wu et al.[289] | Real and imaginary | Bright-field image | U-Net | Expt.: 30,000 pairs | GAN loss |
| | Terbe et al.[290] | Amplitude and phase | Bright-field image | U-Net | Expt.: 3000 unpaired | Cycle-GAN loss |
| | Rivenson et al.[291] | Phase of tissue slices | Stained bright-field image | U-Net | Expt.: >2,000 pairs | GAN loss |
| | Wang et al.[292] | Phase of tissue slices | Stained bright-field or fluorescence image | U-Net | Expt.: 1,000 unpaired | Cycle-GAN loss |
| | Liu et al.[293] | Amplitude and phase of three wavelength | Stained bright-field image | U-Net | Expt.: 8,928 pairs | GAN loss |
| | Nygate et al.[294] | Phase and gradiences of sperm cells | Stained bright-field image | U-Net | Expt.: 1,100 pairs | GAN loss |
| | Guo et al.[295] | Phase, retardance and Orientation | Fluorescence image | 2.5D U-Net | Expt.: 200 full brain sections | $l_1$-norm |
| | Kandel et al.[296,297] | Phase | Fluorescence image | U-Net | Expt.: 30-3,000 pairs | $l_2$-norm |
| | Guo et al.[298] | Phase at different depths | Fluorescence images at different depths | U-Net | Expt.: 200 pairs | $l_2$-norm |
| | Chen et al.[299,300] | Three neighbouring phase | Corresponding central fluorescence image | U-Net and EfficientNet | Expt.: 20 z-stacks | $l_2$-norm |
| | Jo et al.[301] | 3D RI tomogram | 3D fluorescence image | 3D U-Net | Expt.: 1,600 pairs | $l_2$-norm and gradient difference |

## 5.1 Segmentation

Image segmentation, aiming to divide all pixels into different regions of interest, is widely used in biomedical analysis and diagnosis. For un-labeled cells or tissues, the contrast of the bright field intensity is low and thus inefficient to be used for image segmentation. Therefore, segmentation according to the phase distribution of cells or tissues becomes a potentially more efficient way. Given the great success of CNNs in semantic segmentation[302], it seems that we can easily transplant it for phase segmentation, that is, doing segmentation with the phase as input of the neural network (Fig. 27).

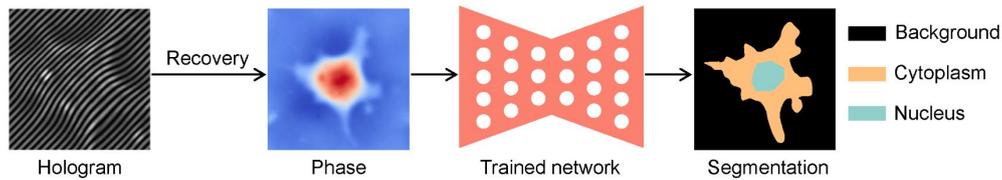

Fig. 27 Description of deep-learning-based segmentation from the phase.

To the best of our knowledge, early in 2013, Yi et al.[303] first proposed to do segmentation from the phase distribution for the red blood cells, although using a non-learning image-processing-based algorithm. To improve the segmentation accuracy in the case of heavily overlapped and multiple touched cells, they first introduced the fully convolutional network (FCN)[302] into phase segmentation[257]. Earlier in the same year, Nguyen et al.[304] used the random forest algorithm to segment prostate cancer tissue from the phase distribution. Ahmadzadeh et al.[258] used the FCN-based phase segmentation to do nucleus extraction for cardiomyocyte characterization. Subsequently, the U-Net was used for phase segmentation in multiple biomedical applications, such as segmentation of the sperm cells' ultrastructure for assisted reproductive technologies[259], SARS-CoV-2 detection[260], cells live-dead assay[261], and cells cycle-stage detection[262]. In addition, other types of neural networks were used for phase segmentation, including the mask R-CNN for cancer screening[263] and the DeepLab-V3+ for cytometric analysis[264].

Further than the phase, the RI from ODT can be used to segment a sample in three dimensions. Lee et al.[265] obtained the 3D shape and position of the organelles by 2D segmentation of the RI tomograms at different depths, which are respectively used for the analysis of the morphological and biochemical parameters of breast cancer cells' nuclei. As a more direct and efficient way, Choi et al.[266] used a 3D U-Net to segment subcellular compartments directly from s single 3D RI tomogram.

5.2 Classification

Similar but different from the segmentation, the classification task is only responsible for giving the overall category of the input sample image, regardless of the specific pixels in the image. For the classification task, the phase provides more information related to the RI and three-dimensional topography of the sample, making it ideal for transparent samples such as cells, tissues, and microplastics[267,305]. Conventional machine learning algorithms first manually extract tens of features from the phase and then do classification with different models. Support vector machine[306], as one of the most popular conventional machine learning strategies, is the most used strategy in phase classification[307–314]. In addition, some researchers used other conventional machine learning strategies, such as k-nearest neighbor[315,316], fully-connected neural networks[317,318], random forest[319,320], and random subspace[321]. More generally, some researchers compared the accuracy of different conventional machine learning strategies in the same application context[317,322–324].

Different from conventional machine learning strategies that require manual feature extraction, deep learning usually takes the phase or its further version directly as input, in which the deep CNNs will automatically perform feature extraction (Fig. 28). This automatic feature extraction strategy tends to achieve higher accuracy, but usually requires a larger number of paired input-label datasets as support. The use of phase as input to deep CNNs for classification was first reported in the work of Jo et al.[267]. They revealed that, for cells like anthrax spores, the accuracy of the neural network using phase as input is higher than that of the neural network using binary morphology image obtained by conventional microscopy as input. Subsequently, this deep-learning-based phase classification method has been used in multiple applications, including assessment of T cell activation state[268], cancer screening[269], classification of sperm cells under different stress conditions[270], prediction of living cells mitosis[271], and classification of different white blood cells[272]. Accuracy in these applications is generally higher than 95% for the binary classification, but cannot achieve comparable accuracy in multi-type classification.

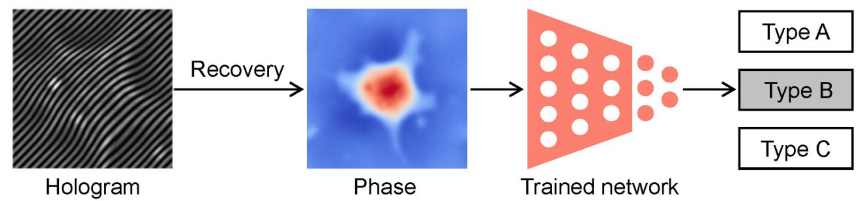

Fig. 28 Description of deep-learning-based classification from the phase.

On the one hand, combining the automatically extracted features of the neural network and the manually extracted features for classification can effectively improve the accuracy, which is because the manually extracted features add the prior of human experts to the classifier[273–275]. For instance, after adding the manual morphological features, the accuracy and area under the curve of healthy and sickle red blood cells classification are improved from 95.08% and 0.9665 to 98.36% and 1.0000[274]. On the other hand, the classification accuracy can also be enhanced by using higher dimensional data of the phase or other data together with the phase as the input of the neural network, such as 3D RI tomogram from the phase[276,277], more phase in temporal dimension[278–280], more phase in wavelength dimension[281,282], and amplitude together with the phase[283–288].

**3D RI tomogram from the phase (Fig. 29a).** Ryu et al.[276] used the 3D RI tomogram as the input of a neural network to classify different types of cells, and achieved an accuracy of 99.6% in the binary classification of lymphoid and myeloid cells, and of 96.7% even in five-type classification of white blood cells. For the multi-type classification, they also used the amplitude or phase of the same sample as input to train and test the same neural network, but only achieved an accuracy of 80.1% and 76.6%, respectively. Afterward, Kim et al.[277] from the same group applied this technology to microbial identification and reached 82.5% accuracy from an individual bacterial cell or cluster for the identification of 19 bacterial species.

**More phase in temporal dimension (Fig. 29b).** Wang et al.[278] used the amplitude and phase from time-lapse holograms as inputs to a pseudo-3D CNN to classify the type of growing bacteria, shortening the detection time by >12 h compared with the environmental-protection-agency-approved methods. Likewise, Liu et al.[279] used the phase from time-lapse holograms as neural network inputs to infer the plaque-forming units probability for each pixel, achieving >90% plaque-forming units detection rate in <20 hours. By contrast, Batuch et al.[280] proposed to use the phase at a specific moment and the corresponding spatiotemporal fluctuation map as the inputs of a neural network to improve the accuracy of cancer cell classification.

**More phase in wavelength dimension (Fig. 29c).** Singla et al.[281] used the amplitude and phase of the red-green-blue color wavelengths as inputs of a neural network, thereby achieving a classification accuracy of 97.7% for healthy and malaria-infected red blood cells, and classification accuracy of 91.2% even for different stages of malaria-infection. Similarly, With the blessing of information from the red-green-blue color holograms, Isil et al.[282] achieved the high-accuracy four-type classification of algae, including accuracy of 94.5%, 96.7%, and 97.6% for *D. tertiolecta*, *Nitzschia*, and *Thalassiosira algae*, respectively.

**Amplitude together with the phase (Fig. 29d).** Lam et al.[284,285] used the amplitude and phase as the inputs of a neural network to do the classification of occluded and/or deformable objects, and achieved accuracy over 95%. With the same strategy, they performed a ten-type classification for biological tissues with an accuracy of 99.6%[286]. Further, Terbe et al.[287] proposed to use a type of volumetric network input by supplementing more amplitude and phase in different defocus distances. They built a more challenging dataset with seven classes by alga in different counts, small particles, and debris. The network with volumetric input outperforms the network with a single amplitude and phase inputs in all cases by approximately 4% accuracy. Similarly, Wu et al.[288] used real and imaginary parts of the complex field as network input to do a six-type classification for bioaerosols, and achieved an accuracy of over 94%.

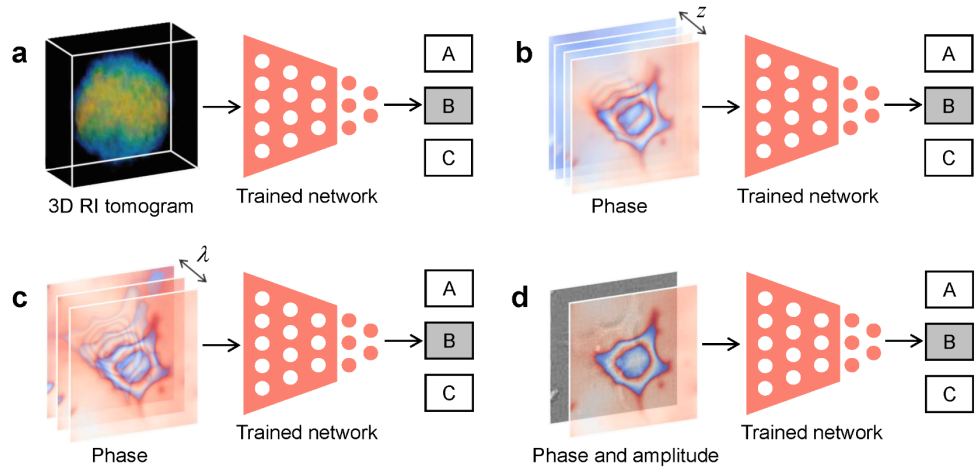

Fig. 29 Description of deep-learning-based classification from higher dimensional data of phase, including (a) 3D RI tomogram, (b) more phase in the temporal dimension, (c) more phase in wavelength dimension, and (d) amplitude together with the phase. **a** Adapted with permission from Ref. [276]. Distributed under Creative Commons (CC BY 4.0) license http://creativecommons.org/licenses/by/4.0/.

In pursuit of extreme speed for real-time classification, some researchers also choose to directly use the raw hologram recorded by the sensor as the input of the neural network to perform the classification tasks[325–329]. Since the information of amplitude and phase are encoded within a hologram, the hologram-trained neural network should achieve satisfactory accuracy with the support of sufficient feature extraction capabilities, which has been proven in practices including molecular diagnostics[325], microplastic pollution assessment[326–328], and neuroblastoma cells classification[329].

5.3 Imaging modal transformation

Let us start this subsection with *image style transfer*, which aims to transfer a given image to another specified style under the premise of retaining the content of this image as much as possible[330,331]. Similarly, for a biological sample, its different parts usually have different RI, different chemical staining properties, or different fluorescent labeling properties, which makes it possible to achieve "image style transfer" from phase recovery/imaging to other different imaging modalities (Fig. 30).

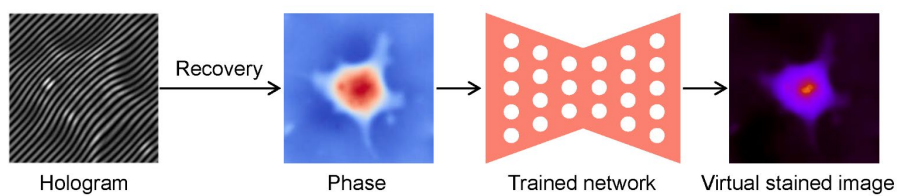

Fig. 30 Description of deep-learning-based imaging modal transformation.

The bright-field images of some color biological samples have sufficient contrast due to their strong absorption of visible light, so for such samples, bright-field imaging can be used as the target imaging modality, in which a neural network is used to transfer the complex image of the sample into its virtual bright-field image. In 2019, Wu *et al.*[289] presented the first implementation of this idea, called bright-field holography, in which a neural network was trained to transfer the back-propagated complex images from a single hologram to their corresponding speckle- and artifact-free bright-field images (Fig. 31a). This type of "bright-field holography" is able to infer a whole 3D volumetric image of a color sample like pollen from its single-snapshot hologram. Further, Terbe *et al.*[290] implemented "bright-field holography" with a cycle-GAN in the case of unpaired datasets.

For most transparent/colorless biological samples, chemical staining enables them to be clearly observed or imaged under bright-field microscopy. This allows the above "bright-field holography" to be used for transparent biological samples as well, which is called virtual staining. Rivenson *et al.*[291] applied this virtual staining technique to the inspection of histologically stained tissue slices and named it PhaseStain, in which a well-trained neural network was used to directly transfer the phase of tissue slices to their bright-field image of virtual staining (Fig. 31b). Using label-free slices of human skin, kidney, and liver tissue, they conducted an experimental demonstration of the efficacy of "PhaseStain" by imaging them with a holographic microscope. The resulting images were compared to those obtained through brightfield microscopy of the same tissue slices that were stained with HandE, Jones' stain, and Masson's trichrome stain, respectively. The reported "PhaseStain" greatly saves time and costs associated with the staining process. Similarly, Wang *et al.*[292] applied the "PhaseStain" in Fourier ptychographic microscopy and adapted it to unpaired dataset with a cycle-GAN. Liu *et al.*[293] used six images of amplitude and phase at three wavelengths as network input to infer the corresponding virtual staining version. In addition to tissue slices, Nygate *et al.*[294] demonstrated the advantages and potential of this deep learning virtual staining approach on a single biological cell like sperm (Fig. 31c). To improve the effectiveness of virtual staining, they used the phase gradients as an additional hand-engineered feature along with the phase as the input of the neural network. In order to assess the effectiveness of virtual staining, they used virtual staining images, phase, phase gradients, and stain-free bright-field images as input data for the five-type classification of sperm, and found that the recall values and F1 scores of virtual staining images were higher than those of other data twice or even four times. This type of single-cell staining approach provides ideal conditions for real-time analysis, such as rapid stain-free imaging flow cytometry.

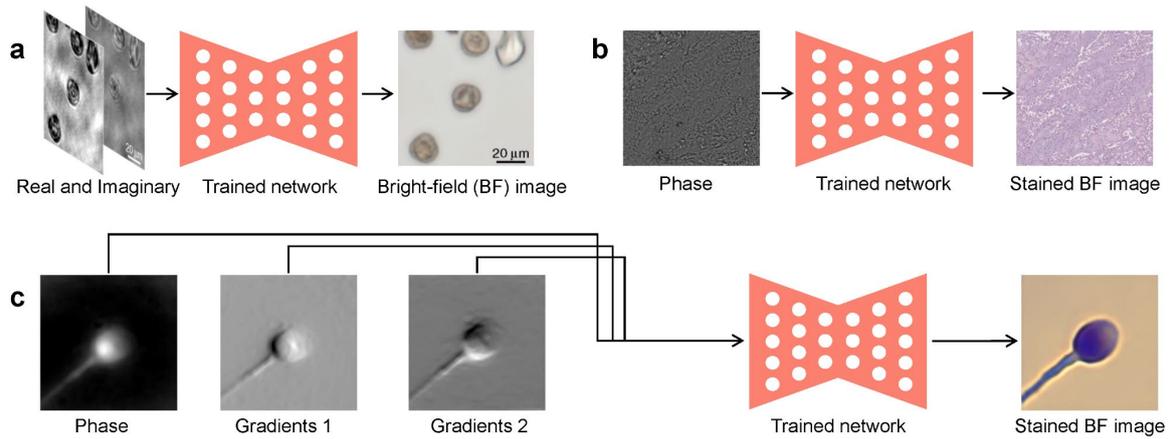

Fig. 31 Description of deep-learning-based virtual staining. **a** Inferring bright-field image from real and imaginary parts. **b** Inferring stained bright-field image from the phase. **c** Inferring stained bright-field image from the phase and its gradients. **a**, **b** Adapted with permission from ref.[289,291]. Distributed under Creative Commons (CC BY 4.0) license http://creativecommons.org/licenses/by/4.0/. **c** Adapted with permission from ref.[294]. Distributed under Creative Commons (CC BY-NC-ND 4.0) license https://creativecommons.org/licenses/by-nc-nd/4.0/.

Apart from imaging color or chemical-stained biological samples with bright-field microscopy, fluorescence microscopy can provide molecular-specific information by imaging fluorescence-labeled biological samples. As a labeled imaging method, fluorescence microscopy has insurmountable disadvantages, including phototoxicity and photobleaching. Guo *et al.*[295] proposed the concept of "transferring the physical-specific information to the molecular-specific information via a trained neural network" (Fig. 32a). Specifically, they used the phase and polarization of cell samples as multi-channel inputs to infer the corresponding fluorescence image, and further demonstrated its performance by imaging the architecture of brain tissue and prediction myelination in slices of a developing human brain. Almost simultaneously, Kandel *et al.*[296] used a neural network to infer the fluorescence-related subcellular specificity from a single phase, which they called phase imaging with computational specificity (Fig. 32b). With these label-free methods, they monitored the growth of both nuclei and cytoplasm for live cells and the arborization process in neural cultures over many days without loss of viability[297]. Guo *et al.*[298] further inferred the fluorescence images from the phase at different depths and performed 3D prediction for mitochondria. The above methods are performed on wide-field fluorescence microscopes, which cannot provide high-resolution 3D fluorescence data for neural networks as ground-truth. Hence, Chen *et al.*[299,300] presented an artificial confocal microscopy consisting of a commercial confocal microscope augmented by a laser scanning gradient light interference microscopy system. To obtain the paired dataset, the artificial confocal microscopy can provide the phase of the samples in the same field of view as the fluorescence channel. With the support of deep learning, their proposed artificial confocal microscopy combines the benefits of non-destructive phase imaging with the depth sectioning and chemical specificity of confocal fluorescence microscopy.

Unlike inferring the fluorescence image from the phase, RI is an absolute and unbiased quantity of biological samples, so a neural network trained with RI as input is naturally applicable to new species. Jo et al.[301] thus built a bridge from ODT to fluorescence imaging via deep learning (Fig. 32c). They trained a neural network with the 3D RI tomogram as input and the corresponding fluorescence image as ground-truth. With the trained neural network, they performed various applications within the endogenous subcellular structures and dynamics profiling of intact living cells at unprecedented scales.

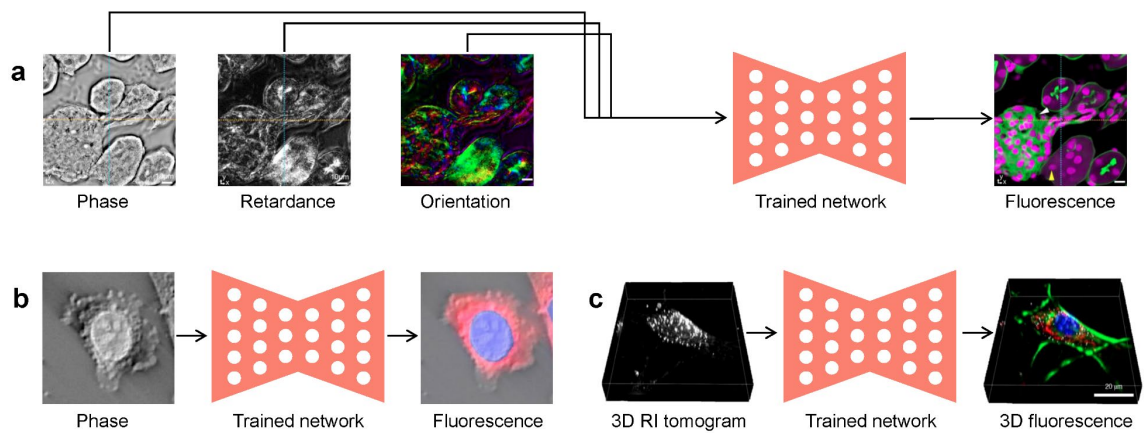

Fig. 32 Description of deep-learning-based label-free virtual fluorescence imaging. **a** Inferring fluorescence image from the phase, retardance, and orientation. **b** Inferring fluorescence image from the phase. **c** Inferring 3D fluorescence image from a 3D RI tomogram. **a**, **b** Adapted with permission from ref.[295,296]. Distributed under Creative Commons (CC BY 4.0) license http://creativecommons.org/licenses/by/4.0/. **c** Adapted with permission from ref.[301]. Springer Nature.

## 6. Conclusion and outlook

The introduction of deep learning provides a data-driven approach to various stages of phase recovery. Based on where they are used, we provided a comprehensive review of how neural networks work in phase recovery. Deep learning can provide pre-processing for phase recovery before it is performed, can be directly used to perform phase recovery, can post-process the initial phase obtained after phase recovery, or can use the recovered phase as input to implement specific applications. Despite the fact that deep learning provides unprecedented efficiency and convenience for phase recovery, there are some common general points to keep in mind when using this learn-based tool.

**Datasets.** For supervised-learning mode, a good paired dataset provides enough rich and high-quality prior knowledge as a guide for neural network training. As one of the most common ways, some researchers choose to collect the intensity image of the real sample through the experimental setup as the input, and calculate the corresponding phase through conventional model-based methods as the ground-truth. Numerical simulations can be a convenient and efficient way to generate datasets for some cases, such as hologram resolution enhancement[51] and phase unwrapping[48]. The paired dataset thus implicitly contains the input-to-label mapping relationship in a large number of specific samples, which determines the upper limit of the ability of the trained neural network. For instance, if the dataset is collected under fixed settings, the trained neural network can only target a fixed device parameter (such as defocus distance, off-axis angle, and wavelength) or a certain class of samples, but cannot adapt to other situations that are not implied in the dataset. Of course, one can ameliorate this by using different settings and different types of samples when collecting datasets, thereby including various cases in the paired training samples, such as adapting to a certain range of defocus distance[95,142], adapting to different aberrations[99,108], adapting to different off-axis angles[103] and adapting to more types of samples[107]. One can use Shannon entropy to quantitatively represent the richness of the amount of information contained in the dataset, which directly affects the generalization ability of the trained neural network[97]. In addition, the spatial frequency content of the training samples in datasets also limits the ability of the trained neural network to resolve fine spatial features, which can be improved to some extent by pre-processing the power spectral density of the training samples[96]. For semi-supervised-learning mode, the cycle-GAN-based method uses an unpaired dataset to train the neural network for learning the mapping relationship between the input domain and the target domain, including phase recovery[104,105,119], noise reduction[184], resolution enhancement[200], and imaging modal transformation[290,292]. As for self-supervised-learning mode, under the guidance of the input-only dataset and the forward physical model, neural networks can learn the inverse process[127,128,131–134].

**Networks and loss functions.** Guided/Driven by the dataset, the neural network is trained to learn the mapping relationship from the input domain to the target domain by minimizing the difference between the actual output and the ground-truth (loss functions). Therefore, the fitting ability of the neural network itself and the perception ability of the loss function determines whether the implicit mapping relationship in the dataset can be well internalized into the neural network. Conventional encoder-decoder-based neural networks have sufficient receptive fields and strong fitting capabilities, but down-sampling operations such as max-pooling lose some high-frequency information. Dilated convolutions can improve the receptive field while retaining more high-frequency information[118]. In addition, convolution in the Fourier frequency domain guarantees a global receptive field, since each pixel in the frequency domain contains contributions from all pixels in the spatial domain[121,122]. In order to make the neural network more focused on different spatial frequency information, one can also use two neural networks to learn the high- and low-frequency bands, respectively, and then use the third neural network to merge them into a full spatial frequency version[144]. Neural architecture search is another potential technology, which automatically searches out the optimal network structure from a large structure space[123]. As the most commonly used loss functions, $l_2$-norm and $l_1$-norm are more responsive to low-frequency information and less sensitive to high-frequency information. That is to say, the low-frequency information in the output of the neural network contributes more to the $l_2$-norm and $l_1$-norm loss functions than the high-frequency information. Therefore, some researchers have been trying to find more efficient loss functions, such as NPCC[96], GAN loss[109,116,117], and default feature perceptual loss of VGG layer[143]. So far, what kind of neural network and loss function is the best choice for phase recovery is still inconclusive.

**Network-only or physics-connect-network (PcN).** Network-only strategy aims to infer the final phase from the raw measured intensity image in an end-to-end fashion using a neural network. It's a one-shot approach, letting the neural network do it all in one go. Neural networks not only need to perform regularization to remove twin-image and self-interference-related spatial artifacts but also undertake the task of free-space light propagation. Therefore, the inference results of the network-only strategy are not satisfactory in some severely ill-posed cases, including weak-light illumination[98] and dense samples[114]. Since free-space light propagation is a well-characterized physical model that can be reproduced and enforced numerically, using numerical propagation in front can relief the burden on the neural network and allow it to focus on learning regularization. In fact, PcN can indeed infer better results than network-only in the above ill-posed cases[98,114]. In another similar scheme, the neural network only performs the task of hologram generation before the phase-shifting algorithm, thus achieving better generalization ability than network-only[62]. In addition, using speckle-correlation processing before the neural network makes the trained neural network suitable for unknown scattering media and target objects[332].

**Interpretability.** In phase recovery, learning-based deep learning techniques usually attempt to automatically learn a specific mapping relationship by optimizing/training neural network parameters with the real-world paired dataset. Deep neural networks usually adopt a multi-layer architecture and contain a large number of trainable parameters (even greater than millions), and are thus capable of learning complicated mapping relationships from datasets. Unlike physics-based algorithms, such network architectures that are general to various tasks often lack interpretability, meaning that it is difficult to discover what the neural network has learned internally and what the role of a particular parameter is by examining the trained parameters. This makes one helpless in practical applications when encountering a failure of neural network inference, in which they can neither analyze why the neural network failed for that sample nor make targeted improvements for the neural network to avoid this failure in subsequent uses. The algorithm unrolling/unfolding technique proposed by Gregor and LeCun gives hope for the interpretability of neural networks[182], in which each iteration of physics-based iterative algorithms is represented as one layer of the neural network. One inference through such a neural network is equivalent to performing a fixed number of iterations of the physics-based iterative algorithm. Usually, physics-based parameters and regularization coefficients are transferred into the unrolled network as trainable parameters. In this way, the trained unrolled network can be interpreted as a physics-based iterative algorithm with a fixed number of iterations. In addition, the unrolled network naturally inherits prior structures and domain knowledge from a physics-based iterative algorithm, and thus its parameters can be efficiently trained with a small dataset.

**Uncertainty.** When actually using a trained neural network to do inference for a tested sample, its ground-truth is usually unknown, which makes it impossible to determine the reliability of the inferred results. To address this, Bayesian CNNs perform phase inference while giving uncertainty maps to describe the confidence measure of each pixel of the inferred result[109,333–335]. This uncertainty comes from both the model itself and the data, called epistemic uncertainty and aleatoric uncertainty, respectively. The network-output uncertainty maps are experimentally verified to be highly consistent with the real error map, which makes it possible to assess the reliability of inferred results in practical applications without any ground-truth[109,335]. In addition to Bayesian neural networks, there are three other uncertainty estimation techniques, including single deterministic methods, ensemble methods, and test time augmentation methods[336].

**From electronic neural networks to optical neural networks.** So far, the artificial neural networks involved in this review mostly run in the hardware architecture with electronics as the physical carrier such as the graphic processing unit, which is approaching its physical limit. Replacing electrons with photons is a potential route to high-speed, parallel and low-power artificial intelligence computing, especially optical neural networks[337,338]. Among them, spatial-structure-based optical neural networks, represented by the all-optical diffractive deep neural network[339], are particularly suitable for image processing. Some examples have initially demonstrated the potential of using optical neural networks for phase recovery[340,341].

There is enormous potential and efficiency in learning-based deep neural networks, while conventional physics-based methods are more reliable. We thus encourage the incorporation of physical models with deep neural networks, especially for those well modeling from the real world, rather than letting the deep neural network perform all tasks as *a black box*. One possible way is to consider the dataset, network structures, and loss functions as much as possible during the training stage to obtain a good pre-trained neural network; in actual use, the pre-trained neural network is used for one-time inference to deal with the situation with high real-time requirements, and the physical model is used to iteratively fine-tune the pre-trained neural network to obtain more accurate results.